\documentclass[fleqn,usenatbib]{mnras}

\usepackage{newtxtext,newtxmath}

\usepackage[T1]{fontenc}

\DeclareRobustCommand{\VAN}[3]{#2}
\let\VANthebibliography\thebibliography
\def\thebibliography{\DeclareRobustCommand{\VAN}[3]{##3}\VANthebibliography}

\usepackage{graphicx}	
\usepackage{amsmath}	

\defcitealias{Standing23}{S23}

\title[Masses of TOI-1338/BEBOP-1]{The EBLM project -- XIII. The absolute dynamical masses of the circumbinary planet host TOI-1338/BEBOP-1, and applications to the study of exoplanet atmospheres.}

\author[D. Sebastian et al.]{
D. Sebastian$^{1}$ $^{\href{https://orcid.org/0000-0002-2214-9258}{\includegraphics[scale=0.5]{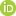}}}$\thanks{E-mail: D.Sebastian.1@bham.ac.uk}, 
A. H.M.J. Triaud$^{1}$ $^{\href{https://orcid.org/0000-0002-5510-8751}{\includegraphics[scale=0.5]{Images/orcid.jpg}}}$,
M. Brogi$^{2,3,4}$ $^{\href{https://orcid.org/0000-0002-7704-0153}{\includegraphics[scale=0.5]{Images/orcid.jpg}}}$,
Thomas A. Baycroft$^{1}$ $^{\href{https://orcid.org/0000-0002-3300-3449}{\includegraphics[scale=0.5]{Images/orcid.jpg}}}$,
Matthew R. Standing $^{5,6}$ $^{\href{https://orcid.org/0000-0002-7608-8905}{\includegraphics[scale=0.5]{Images/orcid.jpg}}}$,
\newauthor
Pierre F.L. Maxted $^{7}$ $^{\href{https://orcid.org/0000-0003-3794-1317}{\includegraphics[scale=0.5]{Images/orcid.jpg}}}$,
David V. Martin $^{8}$ $^{\href{https://orcid.org/0000-0002-7595-6360}{\includegraphics[scale=0.5]{Images/orcid.jpg}}}$,
Lalitha Sairam $^{1,9}$ $^{\href{https://orcid.org/0000-0001-8102-3033}{\includegraphics[scale=0.5]{Images/orcid.jpg}}}$,
Martin B. Nielsen $^{1}$,
\\
$^{1}$ School of Physics \& Astronomy, University of Birmingham, Edgbaston, Birmingham, B15 2TT, UK \\
$^{2}$ Dipartimento di Fisica, Università degli Studi di Torino, via Pietro Giuria 1, I-10125 Torino, Italy\\
$^{3}$ Department of Physics, University of Warwick, Coventry CV4 7AL, UK\\
$^{4}$ INAF-Osservatorio Astrofisico di Torino, Via Osservatorio 20, I-10025 Pino Torinese, Italy\\
$^{5}$ School of Physical Sciences, The Open University, Walton Hall, Milton Keynes. MK7 6AA. UK \\
$^{6}$ European Space Agency (ESA), European Space Astronomy Centre (ESAC), Camino Bajo del Castillo s/n, 28692 Villanueva de la Cañada, Madrid, Spain\\ 
$^{7}$ Astrophysics Group, Keele University, ST5 5BG, UK \\
$^{8}$ Department of Physics \& Astronomy, Tufts University, MA 02155, USA \\
$^{9}$ Institute of Astronomy, University of Cambridge, Madingley road, Cambridge CB3 0HA, UK
\\
}

\date{Accepted. Received; in original form}

\pubyear{2024}

\begin{document}
\label{firstpage}
\pagerange{\pageref{firstpage}--\pageref{lastpage}}
\maketitle

\begin{abstract} 
High-contrast eclipsing binaries with low mass M-dwarf secondaries are precise benchmark stars to build empirical mass-radius relationships for fully convective low-mass ($\rm M_{\star} < 0.35\,M_{\sun}$) dwarf stars. The contributed light of the M-dwarf in such binaries is usually much less than one~per~cent at optical wavelengths. This enables the detection of circumbinary planets from precise radial velocity measurements. 
High-resolution cross-correlation techniques are typically used to detect exoplanet atmospheres. One key aspect of these techniques is the post-processing, which includes the removal of telluric and spectral lines of the host star. We introduce the application of such techniques to optical high-resolution spectra of the circumbinary planet-host TOI-1338/BEBOP-1, turning it effectively into a double-lined eclipsing binary. By using simulations, we further explore the impact of post-processing techniques for high-contrast systems.
We detect the M-dwarf secondary with a significance of 11-$\sigma$ and measure absolute dynamical masses for both components. Compared to previous model-dependent mass measurements, we obtain a four times better precision.  We further find that the post-processing results in negligible systematic impact on the radial velocity precision for TOI-1338/BEBOP-1 with more than $96.6\,$per~cent (1-$\sigma$) of the M-dwarf's signal being conserved.
We show that these methods can be used to robustly measure dynamical masses of high-contrast single-lined binaries providing important benchmark stars for stellar evolution particularly near the bottom of the main sequence. We also demonstrate how to retrieve the phase curve of an exoplanet with high-resolution spectroscopy using our data.
\end{abstract}




\section{Introduction}

Planets orbiting M-dwarfs are at the forefront of exoplanet research, especially if they are temperate rocky exoplanets. Compared to solar-type stars, it is easier to detect planets orbiting M-dwarfs, both in transit as well as in radial velocity surveys \citep[e.g.][]{Triaud2021}. These planets are also particularly favourable candidates to explore the existence and properties of exoplanet atmospheres for mini-Neptunes, super-Earths, and rocky planets \citep{Kaltenegger09,Morley17}. This enables the  study of Earth-sized planets in the habitable zone, which appear to be more abundant for low-mass stars than Sun-like stars \citep[e.g.][]{Dressing13,He2017,Triaud2021}. This is why a number of exoplanet surveys focus on the detection of such planets \citep{Nutzman08,Delrez18,Barclay18,quirrenbach19,Donati20}.

Accurate stellar parameters, such as mass and radius, are essential to characterise these newly discovered worlds and to understand their atmospheres. Obtaining such parameters for isolated M-dwarfs relies on  using stellar models \citep{Dotter08,Baraffe15}, which have not been fully confronted to empirical data. Fully convective stars \citep[$\rm M_{\star} < 0.35\,M_{\sun}$][]{Chabrier97} might be inflated by a few percent compared to models \citep[e.g.][]{Casagrande08,Torres10,Spada13,Kesseli18} which would impact inferences on planetary parameters.


Accurate, precise and absolute parameters can be obtained for stars if they appear in a double-lined eclipsing binary \citep{Hilditch01,Torres10,Triaud2020}. Measurements collected on such systems are used to derive empirical relationships \cite[e.g.][]{Torres10} and confront models such as those by \citet{Dotter08} and \citet{Baraffe15}. However there are very few double-lined eclipsing binaries with component masses $< 0.3~\rm M_\odot$ \citep[e.g.][]{Bender2012,Casewell2018}. This is why another approach was needed.


The Eclipsing Binaries with Low Mass project \citep[EBLM; ][]{triaud13} focuses on high-contrast eclipsing binaries with F,G, \& K-type primary stars, orbited by late-type M-dwarf secondaries, typically with masses $<0.35~\rm M_\odot$. The EBLM project's main goal is to create an empirical mass-radius-luminosity-metallicity relationship. Initially detected by the WASP survey \citep{pollacco06}, a sample of about 200 of such EBLM high-contrast binaries is being analysed \citep[e.g.][]{triaud13,triaud17,vonBoetticher2019}. Typically, EBLM binaries appear as single-lined spectroscopic binaries, which means that masses for the M-dwarf secondaries are inferred by assuming parameters for the solar-like primary star. While this step is usually reliable and leads to precise M-dwarf parameters \citep[e.g.][]{swayne21,swayne23,Sebastian23}, any inaccuracy in the primary star's parameters can lead to systematic biases for the low-mass secondaries \citep{duck23}. As such, single-lined eclipsing binaries are often ignored in the calculation of empirical relationships \citep[e.g.][]{Torres10}. 

As part of the EBLM project, we are now pushing to transform some systems from single-lined to double-lined binaries. Success in that step will allow us to compare the dynamical mass estimates for the M-dwarf secondary to the inferred parameters based on an assumed primary star mass and calibrate results for the EBLM project. In addition, such systems would themselves be included in eclipsing double-lined binary catalogues such as the Detached Eclipsing Binary CATalog \footnote{\url{https://www.astro.keele.ac.uk/jkt/debcat/}} \citep[DEBCAT; ][]{Southworth2015} and used for empirical relationships using dynamically determined masses. We have had recent success with EBLM J0113+31 \citep{Maxted2022}, and this paper is our second double-lined system.

A subset of the EBLM systems are intensively monitored with radial-velocities to search for circumbinary planets as part of the BEBOP survey \citep[Binaries Escorted By Orbiting Planets;][]{martin_2019}. In this paper, we focus on EBLM J0608-59, which is chief amongst those. It is also known as TOI-1338 since the NASA {\it TESS} mission \citep[Transiting Exoplanet Survey Satellite][]{ricker15} identified a transiting circumbinary planet \citep{kostov20}. It is also known as BEBOP-1 since later, the BEBOP project discovered a second, outer planet in that system, using radial-velocities obtained with the HARPS and ESPRESSO spectrographs \citep{Standing23}. The binary itself is a $1.2+0.3~\rm M_\odot$ pair on a mildly eccentric, $14$ day orbital period.

The abundance and quality of high-resolution spectroscopic data make EBLM J0608-59 / TOI-1338 / BEBOP-1 a perfect system to attempt to recover the weak absorption lines of the secondary star. To do that we employ a method typically used to retrieve the emission spectrum of hot Jupiters \citep[e.g.][]{Brogi2012,Birkby2018} using the High-Resolution Cross-Correlation Spectroscopy method \citep[HRCCS;][]{snellen2010}. This has been used for more than a decade to make robust detections of atomic and molecular species and even isotopes within exoplanet atmospheres \citep[e.g.][]{Brogi2012,Line2021,Zhang2021}. One of the main characteristics of this method is the post-processing of the stellar spectra to remove telluric or stellar contributions from the host-star. The reason this method is appropriate is because the flux ratio between a hot Jupiter and its Sun-like host star is similar to that of a very low mass M dwarf orbiting also a Sun-like star. Both hot Jupiters and low mass stars have similar sizes and temperatures making them share a similar parameter space \citep[e.g.][]{Triaud2014,Dransfield2020}.

In this paper (Sections~\ref{HRCCS1} and \ref{detect}), we apply the HRCCS method to optical high-resolution spectra of the circumbinary planet-host binary TOI-1338/BEBOP-1. 
The successful application of these methods provides model-independent, absolute, dynamical masses and, thus our analysis turns what used to be a single-lined planet-host binary into a benchmark double-lined system for stellar models. Furthermore, measuring dynamical masses of planet-host stars means we can also derive dynamical masses for their circumbinary planets.

One advantage we have when applying the HRCCS method to EBLM systems compared to hot-Jupiters is that we can use spectra of isolated M-dwarfs to make a template mask for the cross-correlation. 
In addition TOI-1338/BEBOP-1 received more data than hot Jupiters typically receive for atmospheric investigations.  Section~\ref{measure} is about exploring how the signal of the secondary is retrieved from the data. We analyse in Sections~\ref{SVD_effects} the characteristic effects, what the post- processing might introduce on the detected signal, and the achieved radial velocity precision. We use this to derive dynamical masses of both stars in Section~\ref{masses} and demonstrate in Section~\ref{phasecurve} how the secondary star's spectrum is retrieved at different orbital phases. Applied to exoplanets, this would permit us to construct phase curves at high-resolution and identify the longitude of specific molecules and isotopes.

\section{Data}

High-resolution spectra were obtained using the stabilised spectrographs ESPRESSO \citep{pepe21} at the VLT, and HARPS \citep{Mayor03} at the ESO 3.6\,m telescope at the La Silla observatory, Chile. Thanks to the high stability and resolving power of both spectrographs, these data were successfully used to confirm the existence of the planet BEBOP-1c \citep[][hereafter S23]{Standing23}. Observations were each done with one fibre on the star and a second fibre on the sky, allowing optimal sky background correction. Data reduction was performed by \citetalias{Standing23}, using the ESPRESSO and HARPS data reduction pipelines and provided via private communication.

123 ESPRESSO spectra were taken between 2019 and 2022. We exclude three data points taken during primary eclipses to avoid phases affected by the Rossiter-McLaughlin effect \citep[e.g.][]{Triaud2018} and one point taken during a secondary eclipse, at which point the secondary is not visible. We also exclude 16 spectra identified by \citetalias{Standing23}, who are using a mixture of bisector, FWHM, and student-t test to the radial-velocity measurements to identify outliers. This selection results in 103 ESPRESSO spectra, with a mean signal-to-noise ratio (SNR) of 40 (@550\,nm), which we use in this analysis.

70 HARPS spectra were taken between April 2018 and September 2022 with an average exposure time of 1800\,s and weighted mean SNR of 31.2. Similar to \citetalias{Standing23}, we exclude one spectrum, which appeared to be a wrong pointing as well as two spectra, flagged as outliers, resulting in 67 spectra used in our analysis \citep[The list of excluded spectra, as well as the measured radial velocities can be found in Table 3 and 4 in][]{Standing23}.

Assuming a blackbody approximation, combined with the effective temperatures and radii of both stars (as listed Table\ref{tab:bin_par}), we can derive an average contrast ratio of 0.2~per~cent for the spectral range of ESPRESSO. Thus, the secondary's SNR  is only about 0.09 per spectrum, making it undetectable in individual spectra due to the presence of the primary star. This makes post processing necessary, where we remove strong data features first before searching for the secondary's signal in the residuals.

\section{Post processing} \label{HRCCS1}

For both instruments, we use the one dimensional spectra ({\tt s1d}), provided by the ESO data reduction pipelines, rather than the single spectral orders ({\tt e2ds}). In this way, we use the optimised combination of overlapping spectral orders by the pipeline to work with the highest SNR.

For the post processing, we make use of two facts. First, the secondary's signal is deeply buried in the Poisson noise of the data, produced by the bright primary star. Second, the secondary is moving by several $\rm km\,s^{-1}$ compared to the primary's and instrumental rest frame. This allows us to implement detrending methods, typically used to detect faint atmospheric signals from exoplanets \citep[see][for a review]{Birkby2018}. The method relies on the fact that the lines of the planet host star as well as the telluric contribution are quasi-stationary during planetary transits, and therefore, can be removed as correlated signal from the data without removing the secondary's spectrum. Our case is different to an exoplanet transit observation because the primary stellar spectrum is moving by tens of $\rm km\,s^{-1}$ in the instrument's rest frame, and cannot be assumed to be quasi stationary. We therefore decide to treat the primary's spectrum as a correlated signal and avoid spectral orders with telluric contamination in our analysis. 

The reduced ESPRESSO spectra used in this analysis are automatically wavelength calibrated for vacuum wavelengths only. We use the vacuum to air conversion as described in \cite{Morton00} to correct the spectra first. This will allow us to cross-correlate the spectra to the ESPRESSO line masks at a later stage of the analysis (see Sec. \ref{detect}). Then, we split the 103 ESPRESSO spectra each in 74 chunks of 5990 pixels, which allows us to apply the detrending in a computational feasible way. In the same way, split the 67 HARPS spectra 60 chunks of 5218 pixels. Since the spectra have been provided with an air wavelength calibration no conversion was necessary.

In a next step, we shift each spectrum in the primary's rest frame using linear interpolation according to the radial velocity measurements provided by \citetalias{Standing23}. Each spectrum is then normalised by dividing it through by a polynomial of order six which only fits the continuum regions of each spectrum. The continuum is found automatically by dividing each chunk into five subsets and applying a two stage median clipping to each of them. In a first iteration all data lower than 0.998 and larger than 1.1 of the median value are clipped. In a second iteration a new median of the remaining data is derived and all data lower than 0.998 and larger than 1.02 of the new median are clipped. This process removes stellar lines as well as emission features or hot pixels from the data. Finally, all remaining data which are larger than the new median by at least 0.5-$\sigma$ (the standard deviation of the clipped data)  are kept and used to fit the continuum. 

\subsection{SVD detrending} \label{svd}

The spectral lines of the primary star are removed, by applying a singular value decomposition \cite[SVD, ][]{kalman1996}. This allows an eigenvalue decomposition of an arbitrary matrix $A$ without the need for deriving the matrix product of $A^{\rm T}A$ as it is the case for principal component analyses. Similar to \cite{deKok13}, we use the spectral array for each chunk as input matrix $A$. Its size is, therefore, defined by the number of pixels in each chunk as well as by the number of observed spectra. The SVD decomposes $A$ as,
\begin{equation}
A = U S V^{\rm T},
\end{equation}
with $V$ and $U$ being orthogonal matrices and $S$ is a diagonal matrix. We use the python module {\tt scipy.linalg.svd} for this decomposition, which returns $V$, $U$, as well as the diagonal elements of $S$ - the eigenvalues - sorted in non-increasing order. Furthermore, \cite{kalman1996} shows that $A$ can be written as an outer product expansion,
\begin{equation}
A = \sum_{i=1}^{R} s_i u_{i} v_{i}^{\rm T},
\end{equation}
with $R$ - the matrix rank - being defined as the smallest dimension of $A$, thus the number of observed spectra in our case. $s_{i}$ are the eigenvalues, $u_{i}$ the columns of $U$, and $v_{i}^{\rm T}$ the rows of $V$. The power of this method is that the first components of this decomposition (with largest eigenvalues) represent features which are strongly correlated to the rows and columns of the spectral matrix, hence the primary's spectrum. Similar to \cite{deKok13}, we now reconstruct the spectral matrix $A'$, excluding the first components as, 
\begin{equation}\label{svd_eq}
A' = \sum_{i=k+1}^{R} s_i u_{i} v_{i}^{\rm T},
\end{equation}
with $k$ being the number of excluded components. A careful selection of $k$ allows us to remove the primary's spectrum. Optimally, $A'$ contains now white noise as well as the secondary's signal.

Typically, the primary's signal is not perfectly correlated with the array axes, resulting in the need to exclude a higher number of components. According to equation~\ref{svd_eq} the allowed number of excluded components can be selected between $k=1$ (only strongest correlation is removed) and $k=R-1$ (only weakest correlation is left). Higher $k$ will result in decreased noise of the residual array, but will also continue to degrade the remaining signal of the secondary. This effect has been observed and analysed for different detrending approaches for exoplanet observations, with the number of removed components typically being optimised as a function of the signal significance \citep[e.g. see][for a discussion]{Cheverall23}. Nevertheless, up to now there is no agreed `best practice' when it comes to selecting the optimal $k$, as it seems to depend on the analysed data-set. 

Different from the methods above, we propose to use the `effective rank' \citep{roy07} to select $k$. It is based on the entropy of the eigenvalues and, thus, depends on the spectral array itself, rather than on the secondary's signal. It can be understood as a measure of correlated signal, and returns the rank where the resulting matrix $A'$ is statistically uncorrelated to the array axes. Therefore the primary's signal is fully removed. For a matrix $A$ the effective rank ($k_{\rm eff}$) can be derived as, 
\begin{equation}
k_{\rm eff}(A) = \exp{H}, 
\end{equation}
with the entropy,
\begin{equation}
H = - \sum_{i=1}^{R} p_{i} \log p_{i}, 
\end{equation}
and the normalised eigenvalues,
\begin{equation}
p_{i} = \frac{s_i}{\sum_{i=1}^{R}\big|s_i\big|}.
\end{equation}

Before we apply the SVD, we first remove outliers, which would otherwise affect the SVD correlation for the affected rows or columns. To achieve this effectively, we make use of the spectra being normalised and aligned to the primary's rest frame. We first average all spectra and then remove the averaged spectrum from the whole array. We apply a 3-$\sigma$ clipping to the residual array and use the outlier positions to mask the spectral array.

\begin{figure}
	\includegraphics[width=\linewidth]{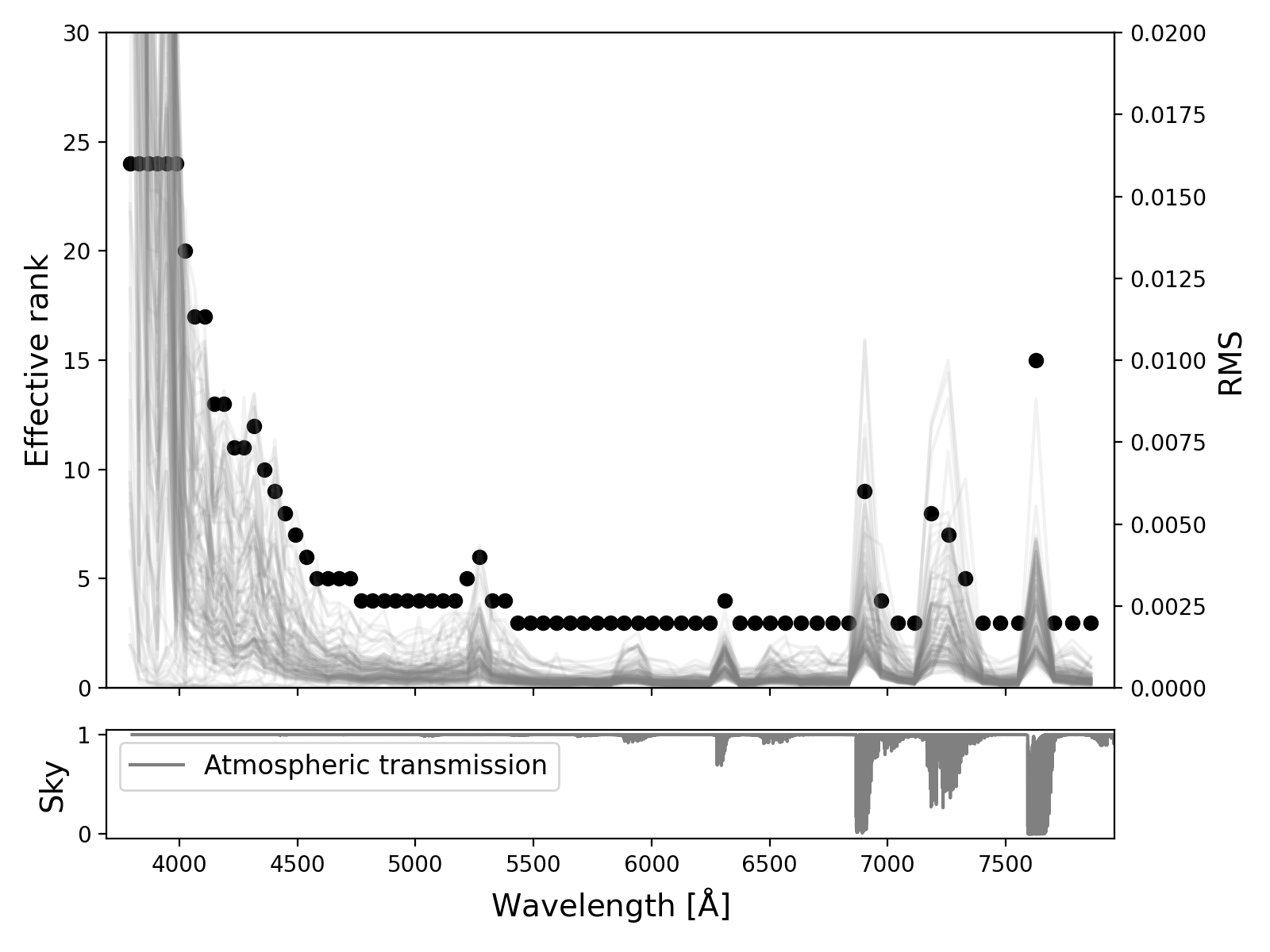}
    \caption{Upper panel: `Auto' SVD detrending for ESPRESSO data. Black dots: Effective rank, Grey lines, RMS of the residual arrays as a function of wavelength. Lower panel: Atmospheric transmission at Paranal Observatory. Wavelength areas with large RMS in the upper panel, match well with strong telluric lines, which are less correlated in the primary's rest frame.}
    \label{fig:eff_rank}
\end{figure}

We then apply the SVD to the normalised array by selecting for each chunk $k$, automatically using $k_{\rm eff}$. In this automatic process, we limit $k$ to $\le 24$ and $\le 12$ for ESPRESSO and HARPS data respectively, to avoid degradation of the secondary signal for noisy data. The median values for $k$ are $k= 4$ and $k=5$ for ESPRESSO and HARPS spectra respectively. Fig.~\ref{fig:eff_rank} shows the effective rank for the ESPRESSO spectra versus the mean wavelength of each chunk, compared to the RMS of the residual array $A'$. Noisy chunks are dominant for the bluest wavelengths due to the decreased detector efficiency as well as the gap between the blue and the red camera at about 627\,nm. We use the ESO {\tt SkyCalc}\footnote{\hyperlink{https://www.eso.org/observing/etc/bin/gen/form?INS.MODE=swspectr+INS.NAME=SKYCALC}{ESO SkyCalc web tool}} web tool to derive the typical atmospheric transmission for ESPRESSO spectra, based on the Cerro Paranal Sky Model \citep{Noll12,Jones13}. Noisy chunks in the red part of the spectra match well to telluric apsorption lines. This is because the telluric lines in these chunks are not correlated to the primary's rest frame, hence the SVD is less effective, which is by design reflected by the effective rank. A similar graph for HARPS data can be found in the appendix (Fig.~\ref{fig:eff_rank_HARPS}). We further explore in Section~\ref{SVD_effects} the effects that the selection of $k_{\rm eff}$ - as well as the different choices of $k$ - for the SVD detrending have on the detected signal of the secondary.

Finally, we apply a 10-$\sigma$ clipping to mask any outliers that might still be in the residual data and finally re-normalise the residual array. We add these outliers to the mask, derived before the SVD detrending.


 
\section{Secondary detection} \label{detect}

To detect the signal of the faint secondary star, we make use of the K-focusing process, a method used for more than a decade to detect the signal of atmospheric atomic and molecular species in exoplanet atmospheres \citep[e.g.][]{snellen2010,Birkby2018,Sebastian23b}. This method is based on a signal enhancement of the secondary's spectrum by first deriving its cross-correlation function \citep[CCF,][]{baranne96,Pepe02} with a suitable template spectrum or line mask, and second by combining the CCFs of all observations within the rest frame of the secondary.

Here we make use of the orbital parameters of the primary, which are known to a high precision. In case of Keplerian two body motion, the secondary shares the same period ($P$), time of periastron ($T_{\rm 0,peri}$), and eccentricity ($e$). The argument of periastron of the secondary ($\omega_{2}$) can be derived from the primary by $\omega_{2}=\omega - \pi$. Thus, the radial velocity of the secondary can be expressed by
\begin{equation} \label{reflex}
    V_{\rm r,2} = K_{2} [\cos(\nu + \omega_{2}) + e \cos(\omega_{2})],
\end{equation}
with $K_{2}$ the secondary's semi-amplitude, and $\nu$ the true anomaly at the time of mid-exposure, determined from $P$ and $T_{\rm 0,peri}$. This assumes that other significant reflex motions have been removed from the data. The primary has been observed for more than two years. Except for the $\sim 5\,{\rm m s^{-1}}$ semi-amplitude of planet c \citepalias{Standing23}, no trend has been measured for this binary. No significant change of $\omega_{2}$ has been observed either. We do not correct for the planet's semi amplitude, as its reflex motion is too small to significantly alter the secondary's orbit.

\begin{figure}
	\includegraphics[width=\linewidth]{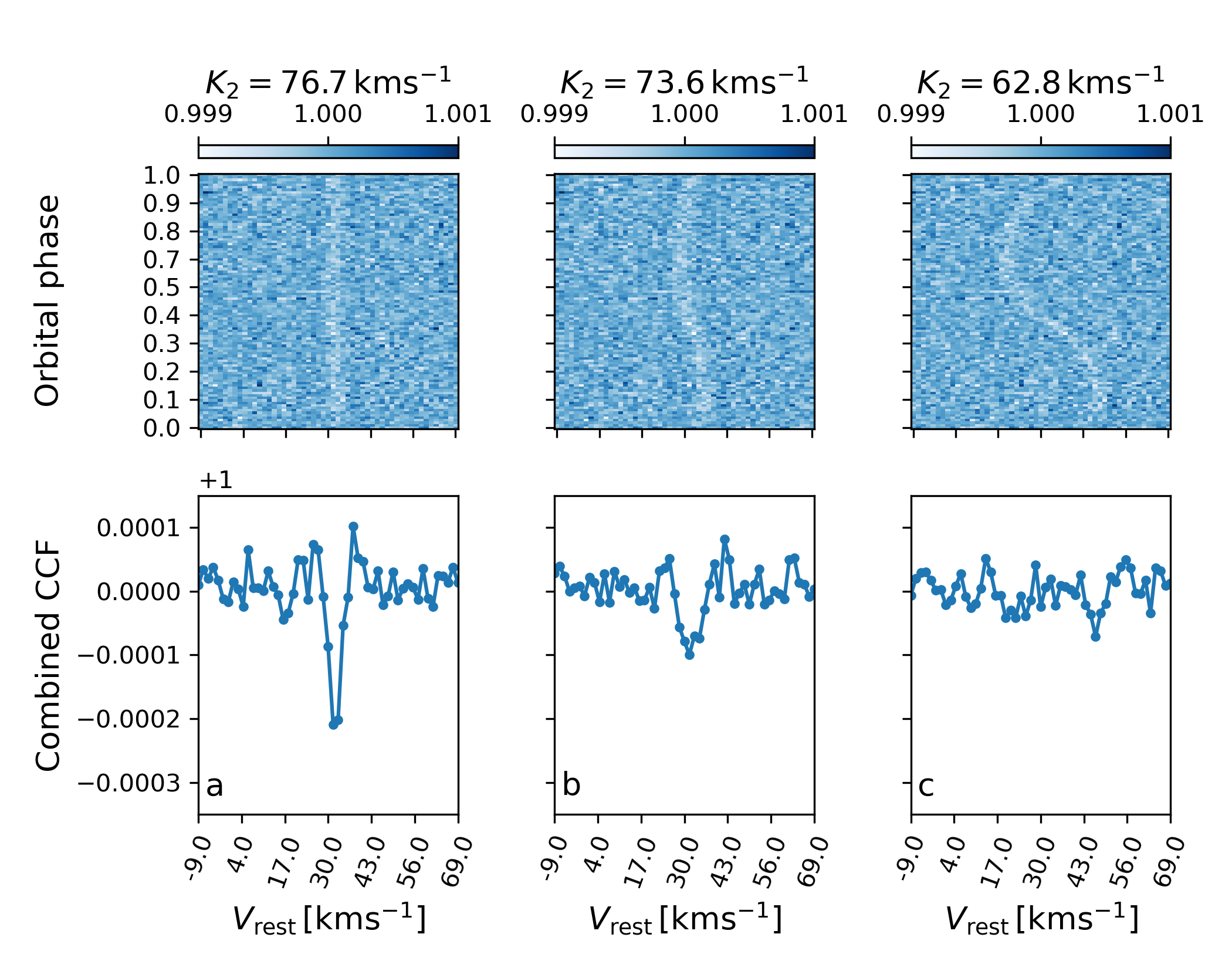}
    \caption{Cross-correlation functions (CCFs) of ESPRESSO spectra with an M2 line mask after the SVD detrending. Upper panel: Individual CCFs for each observation. The shape of the secondary's trail is clearly visible. The secondary is moved into the rest frame using different semi-amplitudes from (a) close to the secondary's semi-amplitude to (c) about 15\,${\rm km s^{-1}}$ smaller. Lower panel: combined CCFs from upper panels. The CCF shape becomes wider from a - c, resulting in weaker absorption structures from different parts of the orbit.}
    \label{fig:sec_det}
\end{figure}

The only unknown parameter of the secondary's orbit is the semi-amplitude $K_{2}$.
To measure its semi-amplitude, we first define a range between $55 - 100\,\rm km\,s^{-1}$ in steps of $1.5\,\rm km\,s^{-1}$. This fully covers the expected semi-amplitude ($K_{\rm 2,exp}$) and its uncertainty, derived from \citetalias{Standing23} as $K_{\rm 1} \times M_1/M_{2} = 77.8\pm5.6\,\rm km\,s^{-1}$.

For a specific $K_{2}$ from this range, we derive the CCF for each individual - post processed - spectrum, creating a two dimensional CCF. This is done by first interpolating the masked spectra to have a uniform sampling of $500\,{\rm m\,s^{-1}}$, which is close to the pixel resolution of both the HARPS and ESPRESSO one-dimensional spectra. Then, we cross-correlate them with an M2-dwarf line mask, shifted to the secondary's rest frame velocity ($V_{\rm rest}$). We use the line masks, which are available in the ESPRESSO and HARPS data reduction pipelines\footnote{The ESPRESSO pipeline has been publicly released on \hyperlink{https://www.eso.org/sci/software/pipelines/espresso/espresso-pipe-recipes.html}{https://www.eso.org}}. These are optimised for high-precision radial-velocity measurements with the respective instruments. This line mask has also been optimised to exclude spectral regions with telluric absorption. We thus, automatically exclude chunks with telluric contributions from the analysis. \citetalias{Standing23} report the systemic velocity $V_{\rm sys}=30.75\,{\rm km\,s^{-1}}$. To include this velocity, we sample the cross-correlation function (CCF) for a velocity range between $-10\,{\rm km\,s^{-1}}$ and $70\,{\rm km\,s^{-1}}$ with a spacing of $1.5\,{\rm km\,s^{-1}}$. 

In Fig.~\ref{fig:sec_det}, we show three of these two-dimensional CCFs derived from ESPRESSO spectra for a $K_{2}$ of (a) $76.3\,{\rm km s^{-1}}$, (b) $73.6\,{\rm km s^{-1}}$, and (c) $62.8\,{\rm km s^{-1}}$. We sort the spectra in order of the orbital phase, which allows to clearly reveal the secondary's trail, which aligns vertically in its rest frame, for a $K_{2}$ close to $K_{\rm 2,exp}$.

\begin{figure}
	\includegraphics[width=\linewidth]{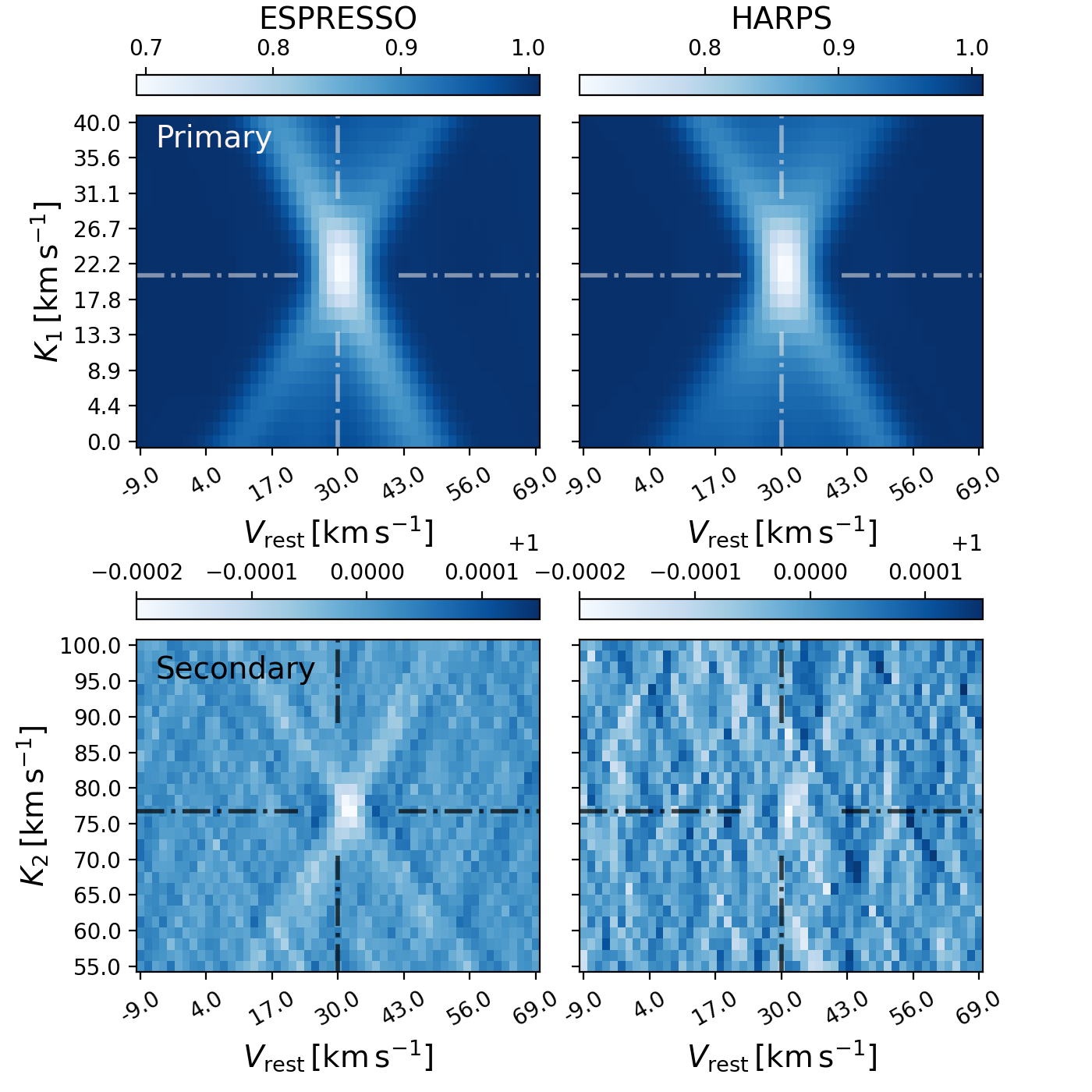}
    \caption{Cross-correlation maps, derived from K-focusing, for ESPRESSO and HARPS data. Upper panel: Primary component. White dashed lines mark the expected semi amplitude and systemic velocity of TOI-1338/BEBOP-1. Lower panel: Detection of the secondary component. Left: 11-$\sigma$ detection for ESPRESSO, Right: 3-$\sigma$ detection for HARPS. Dark dashed lines mark the expected semi amplitude of the secondary as well as the expected systemic velocity measured from the primary.}
    \label{fig:sec_cross}
\end{figure}

We then normalise all individual CCFs for a specific $K_{2}$ by their median, and combine them using a mean function, weighted by the signal-to-noise ratio of the individual observations. This results in a combined CCF for each spectral chunk. Afterwards, we combine the CCFs for all spectral chunks, taking the - SNR weighted - standard deviation of each chunk into account. In this way, we ensure that - despite prior normalisation -  changing data quality and residual noise from telluric lines is effectively taken into account. The lower panels of Fig.~\ref{fig:sec_det} show these combined CCFs visualising the idea of the K-focusing process: The combined CCF signal reaches its largest contrast, the closer $K_{2}$ matches the true semi-amplitude of the secondary. The lower panel in Fig.~\ref{fig:sec_cross} shows the resulting CCF map of combined CCFs for all sampled semi-amplitudes in the $ K_{2} - V_{\rm rest}$ plane for ESPRESSO and HARPS data respectively. The typical shape from the K-focusing process is clearly visible for ESPRESSO data. For the HARPS data, the signal of largest CCF contrast can be clearly seen at a similar position, compared to the ESPRESSO data. Nevertheless, most of its shape - like for typical exoplanet atmospheric detections - is hidden within the noise.
We can apply the same method to the normalised spectra, which have not been post-processed. This allows us to measure independently the semi-amplitude of the primary star using the same method. We derive the CCFs using the same velocity range, we use for the secondary, and sample the semi-amplitude $K_{\rm 1}$ between $0 - 40\,\rm km\,s^{-1}$ in steps of $1.5\,\rm km\,s^{-1}$. The primary's CCF maps are shown for both ESPRESSO and HARPS data in the upper panels of Fig.~\ref{fig:sec_cross}. The primary's semi-amplitude and systemic velocity do visually match well with the maximum CCF contrast. We note an apparent offset of the secondary's rest frame velocity to this systemic velocity in both detections, which we investigate in the following section.



\section{Measuring the signal position} \label{measure}

To measure the signal position, we use the {\tt Saltire} model, which we describe in detail in \cite{Sebastian23b}. The model simulates the CCF signal of the secondary as a one-dimensional double-Gaussian function which undergoes the K-focusing process similarly to the observed spectra. In this way, we create a simulated CCF map which can be fitted to the measured CCF map to derive the best-fitting parameters for $K_{2}$ and $V_{\rm rest}$. The double Gaussian is designed to fit the side-lobes, typically observed for CCFs of M-dwarf spectra \citep[e.g.][]{Bourrier18} . Its five parameters are the mean height ($h$) of the CCF outside the signal, the standard deviations ($\sigma_1$ and $\sigma_2$) of the two Gaussian functions with respective intensities $A_1,\,A_2$ at velocity ($\mu $), the quotient $\Delta = A_2/A_1$, as well as the sum $\Sigma = A_1+A_2$ of both intensities. By selecting an identical $\mu $, but opposite signs for the two intensities, we effectively fit symmetric side-lobes, centred at the same velocity.

Due to the K-focusing, this two dimensional model includes the actual phase coverage of the data, which has shown to result in up to 10 times more accurate measurements for $K_{2}$, \citep{Sebastian23b} compared to the more traditional approach, which is typically a one-dimensional Gaussian fit at the rest velocity of maximum CCF contrast.

The model allows to run a least-squares minimisation, based on {\tt lmfit} \citep{Newville16} as well as to sample the posterior probability distribution, using the Markov chain Monte Carlo (MCMC) code \texttt{emcee} \citep{Foreman-Mackey13}. We first use the least-squares fit with starting parameters, for $K_{2}$, $V_{\rm rest}$, and CCF parameters listed in Table\,\ref{tab:fit_model} with wide uniform priors. We then use the resulting best-fit parameters as input for the MCMC sampling. We sample with 4\,000 calls for 42 parallel chains, rejecting the burn-in samples (the first 1,500 samples) of each walker and thin the remaining samples by a factor 5. This results in a posterior distribution of 21,000 samples. We derive the detection significance as the quotient of $\Sigma/\sigma_{\rm jit}$, which is basically the ratio between the measured CCF contrast and the noise term from the MCMC. Due to the low detection significance of the HARPS CCF map, one parallel chain did not converge and was removed from the sample, resulting in 20,500 samples in the final distribution. We finally repeat the same procedure for the CCF maps of the primary.
In Table~\ref{tab:fit_model}, we report the resulting parameters as the 50th percentile, and errors by averaging the 16/84 percentiles of the posterior distributions. 

\begin{table*}
	\centering
	\caption{50th percentile of the MCMC parameter samples, using the {\tt Saltire} MCMC model for the secondary's and primary's CCF maps. The errors represent the average 16/84 percentiles of each distribution, which are errors from the fit. Systematic errors are discussed in Sec.~\ref{uncs}.}
	\label{tab:fit_model}
	\begin{tabular}{lcccccc}
    \hline
    Parameters & \multicolumn{2}{c}{ESPRESSO} & \multicolumn{2}{c}{HARPS} & \multicolumn{2}{c}{Input fit parameters}\\
                    & Primary & Secondary & Primary & Secondary & Primary & Secondary\\\hline
    $K\,[\rm km\,s^{-1}]$   & $21.6037\pm0.0046$ & $77.161\pm0.078$ & $21.6151\pm0.0046$ & $77.55\pm0.43$ & 21.61 & 77.83 \\
    $V_{\rm rest}\,[\rm km\,s^{-1}]$   & $30.662\pm0.003$ & $32.0584\pm0.0617$ & $30.749\pm0.003$ & $32.41\pm0.27$ & 32 & 32 \\
    $A_{1} + A_{2}$   & $-0.31840\pm0.00023$ & -(2.2925$\pm$0.075)e-04 & $-0.30793\pm0.00024$& -(1.43$\pm$0.24)e-04 & -0.3 & -3e-04 \\
    $A_{2} / A_{1}$   & $-0.15\pm0.16$ & $-0.446388\pm0.040783$ & $-0.147079\pm0.167959$ & $-0.144\pm0.091$ & -0.1 & -0.4 \\
    $\sigma_1\,[\rm km\,s^{-1}]$ &$3.6764\pm0.0085$& $2.334\pm0.088$ & $3.7020\pm0.0174$& $1.66\pm0.29$ & 2.4 & 2.4 \\
    $\sigma_2\,[\rm km\,s^{-1}]$ &$3.676\pm0.065$& $4.170\pm0.218$ & $3.6904\pm0.1320$ & $17.0\pm11.3$ &4.6& 4.6 \\
    $h$                   &1+ (5.000$\pm$0.036)e-03& 1+(8.18$\pm$0.73)e-06 & 1+ (5.801$\pm$0.035)e-03 & 1- (2.50$\pm$6.99)e-06 & 1 & 1\\ 
    $\sigma_{\rm jit}$    &(9.9947$\pm$0.0062)e-04 & (2.138$\pm$00.037)e-05 & (9.9666$\pm$0.0399)e-04& (4.993$\pm$0.085)e-05 & --& --\\
    SNR & 309.0 & 10.7 & 318.6 & 2.9 & --& --\\
    \hline
    \end{tabular}
    \\
\end{table*}
  
\begin{figure*}
	\includegraphics[width=\textwidth]{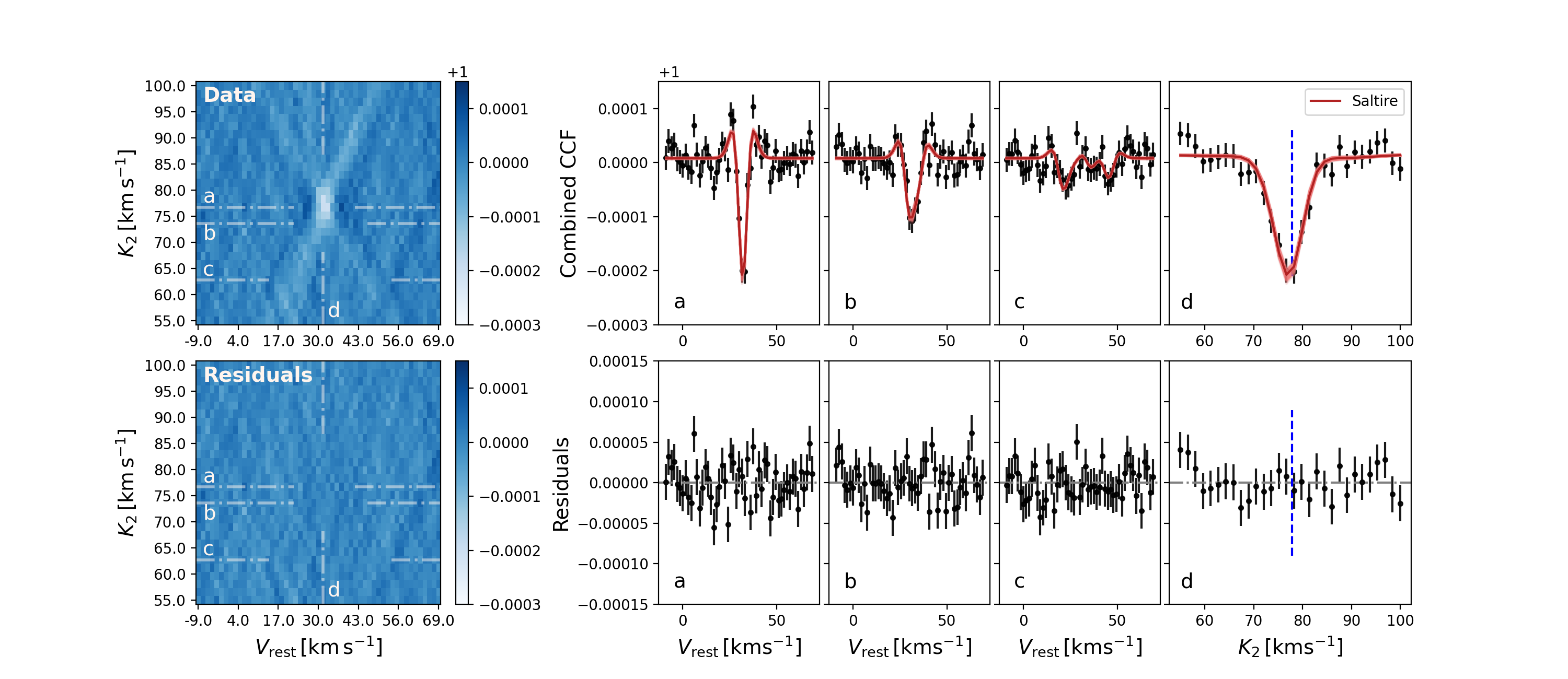}
    \caption{{\tt Saltire} model fit to ESPRESSO detection of the secondary. Upper panels; Left: CCF map. Dashed lines and labels indicate positions of slices through the map, shown on the right. Right: Slices through the CCF map, for different semi-amplitudes (a,b,c, similar to Fig.~\ref{fig:sec_det}), and for the rest velocity at maximum CCF contrast (d); Black dots: CCF map data, Red line: best fitting {\tt Saltire} model. Blue dashed line: Expected semi-amplitude of the secondary from SB1 analysis.
    Lower panels; Left: Residual CCF map. Dashed lines and labels are identical to upper panels. Right: slices through the residuals at the same positions as above. Error bars are derived from 2D {\tt Saltire} fit to the CCF map.}
    \label{fig:ESPRESSO_fit}
\end{figure*}

\begin{figure}
	\includegraphics[width=\linewidth]{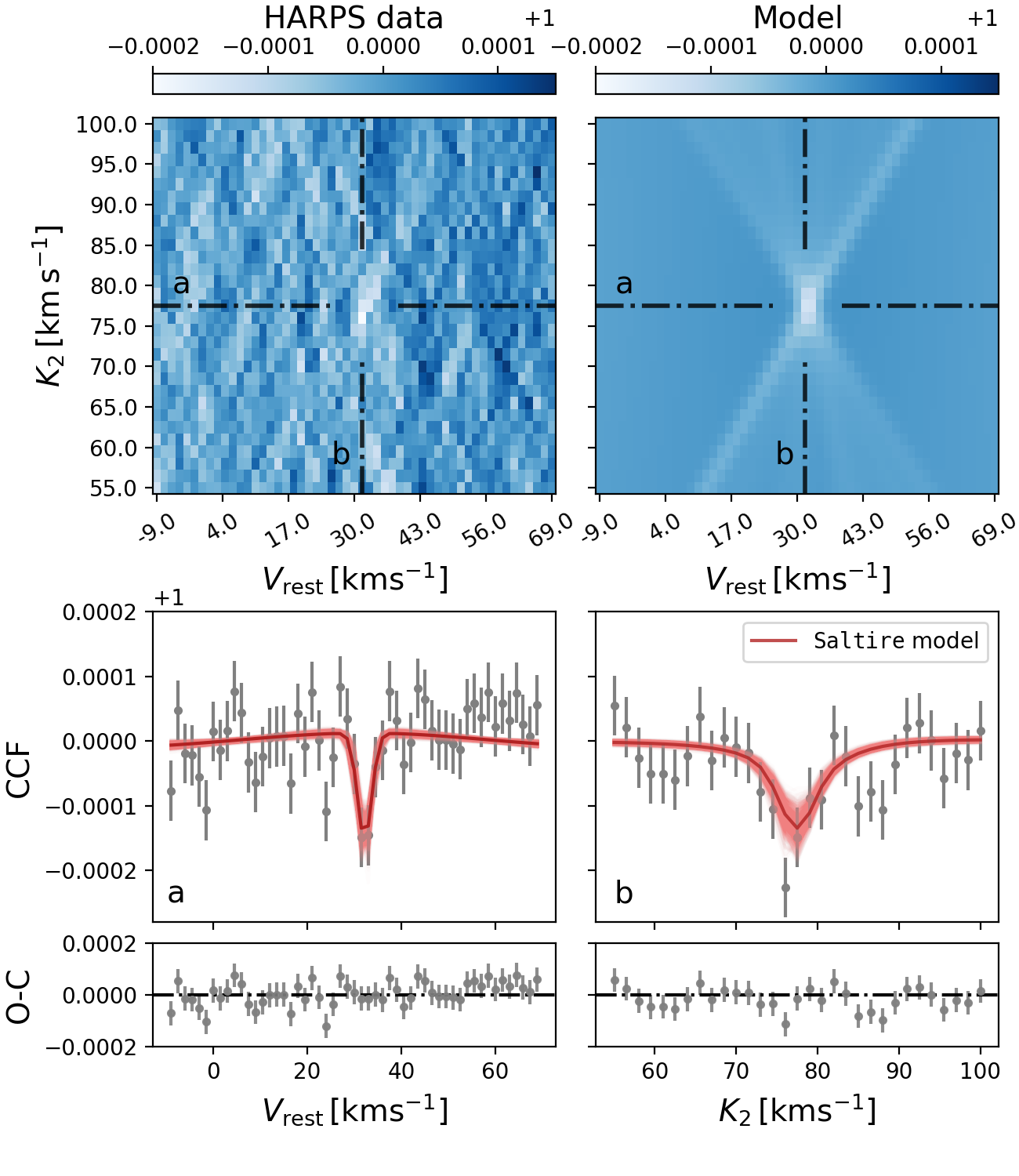}
    \caption{Upper panel: Cross-correlation map and {\tt Saltire} model for HARPS data. Black dotted lines show the positions of slices (a,b), displayed in the lower panel. Lower panel: Slice through CCF map at maximum significance. Red line: best-fitting {\tt Saltire} model, red shaded area: $1\sigma$ Uncertainties from the MCMC. Error bars represent the MCMC jitter term ($\sigma_{\rm jit}$) from the two dimensional fit.}
    \label{fig:HARPS_fit}
\end{figure}

Fig.~\ref{fig:ESPRESSO_fit} shows the best-fitting {\tt Saltire} model for the ESPRESSO data. For clarity, we add slices though the CCF map for both, data and model at constant semi-amplitudes, which are identical to the CCF maps (a,b, \& c), shown in Fig.~\ref{fig:sec_det}. It is clearly visible, that the {\tt Saltire} model is able to represent the shape of the CCF signal at maximum contrast, but also for semi-amplitudes more than $10\,\rm km\,s^{-1}$ away from the semi-amplitude of maximum CCF contrast. Thus, the model is able to fit the whole shape of the signal, making the resulting parameters less biased from local noise structures, close to the signal of maximum contrast. We also show a slice for constant rest velocity at maximum CCF contrast (d). The shape of this slice is not a Gaussian, but solely defined by the K-focusing process \citep[see ][]{Sebastian23b}. We highlight $K_{\rm 2,exp}$, which appears to be $0.7\,{\rm km\,s^{-1}}$ larger, compared to the best-fitting semi-amplitude returned by {\tt Saltire}. We note that {\tt Saltire} performs a two-dimensional fit, thus the error bars, shown in the one-dimensional slices represent the RMS of the two-dimensional fit.
Fig.~\ref{fig:HARPS_fit} shows the best-fitting model for the HARPS data. Clearly the detection is dominated by noise structures due to the lower signal-to-noise of the spectra, which is also reflected in the larger errors. We derive the detection significance as the quotient of the posterior sample of the CCF contrast ($A_1+A_2$) and the RMS of the residual CCF map, which results in a 2.9-${\sigma}$ signal, thus, just at the limit of a detection. Nevertheless, we detect the secondary's signal at a position consistent within the uncertainties with the value of $K_{2}$ measured from ESPRESSO data.

\subsection{Systematic uncertainties} \label{uncs}

Uncertainties returned by {\tt Saltire} represent the precision for the fitted CCF map. Systematic uncertainties from correlated noise, introduced during the creation of the CCF map are not taken into account \citep{Sebastian23b}. For low-significance detections of exoplanet atmospheres, correlated noise can mimic molecule signals, which is why bootstrapping methods have been discussed to verify such detections \citep{Hoeijmakers2020,Borsato2023}. These methods perform the K-focusing process on partial samples, randomly selected from the observations. Assuming white noise, partial samples will each contain a different set of correlated noise. Assuming a constant signal from the secondary component, we can actually disentangle the noise from the signal. Thus, uncertainties accounting for correlated noise can be derived by fitting CCF maps from partial samples.

We make use of this principle by using the high detection significance of the primary's for HARPS and ESPRESSO data, as well as of the secondary detection in the ESPRESSO data set, to characterise systematic uncertainties in the data. For the ESPRESSO data, we divide the data set in four parts - covering different observing seasons - commencing at MJD = 58,736.377, 59,207.187, 59,434.386, \& 59,555.105 with 19, 25, 29, and 30 spectra each. We split the HARPS data in four parts commencing at MJD = 58,436.321, 58,698.435, 59,163.205, \& 59432.415 with 9, 12, and 20, and 25 spectra each. Here we exclude the very first observing season commencing at MJD = 58,216.029 with only one single HARPS spectrum taken.

\begin{figure}
	\includegraphics[width=0.45\linewidth]{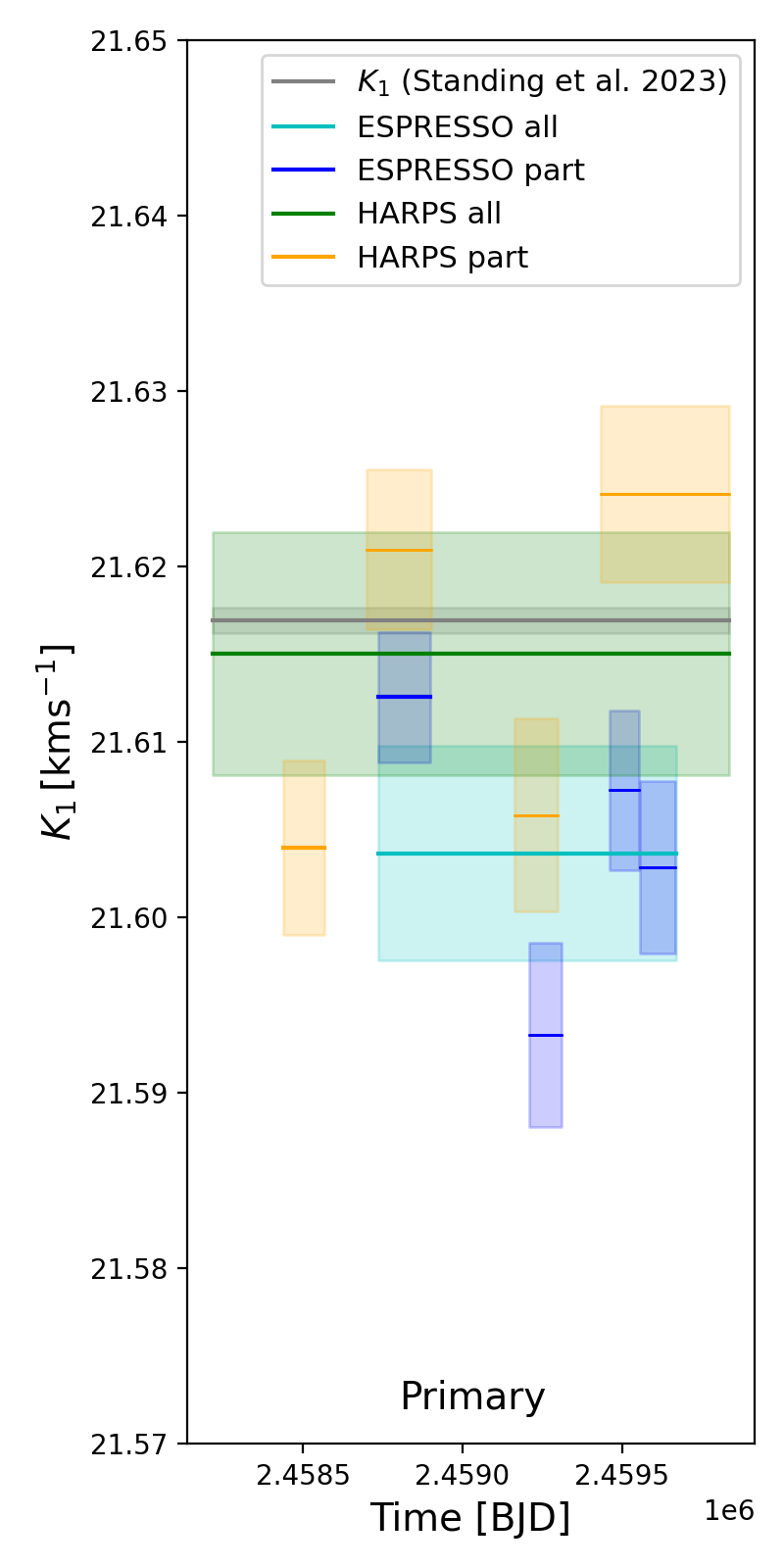}
        \includegraphics[width=0.45\linewidth]{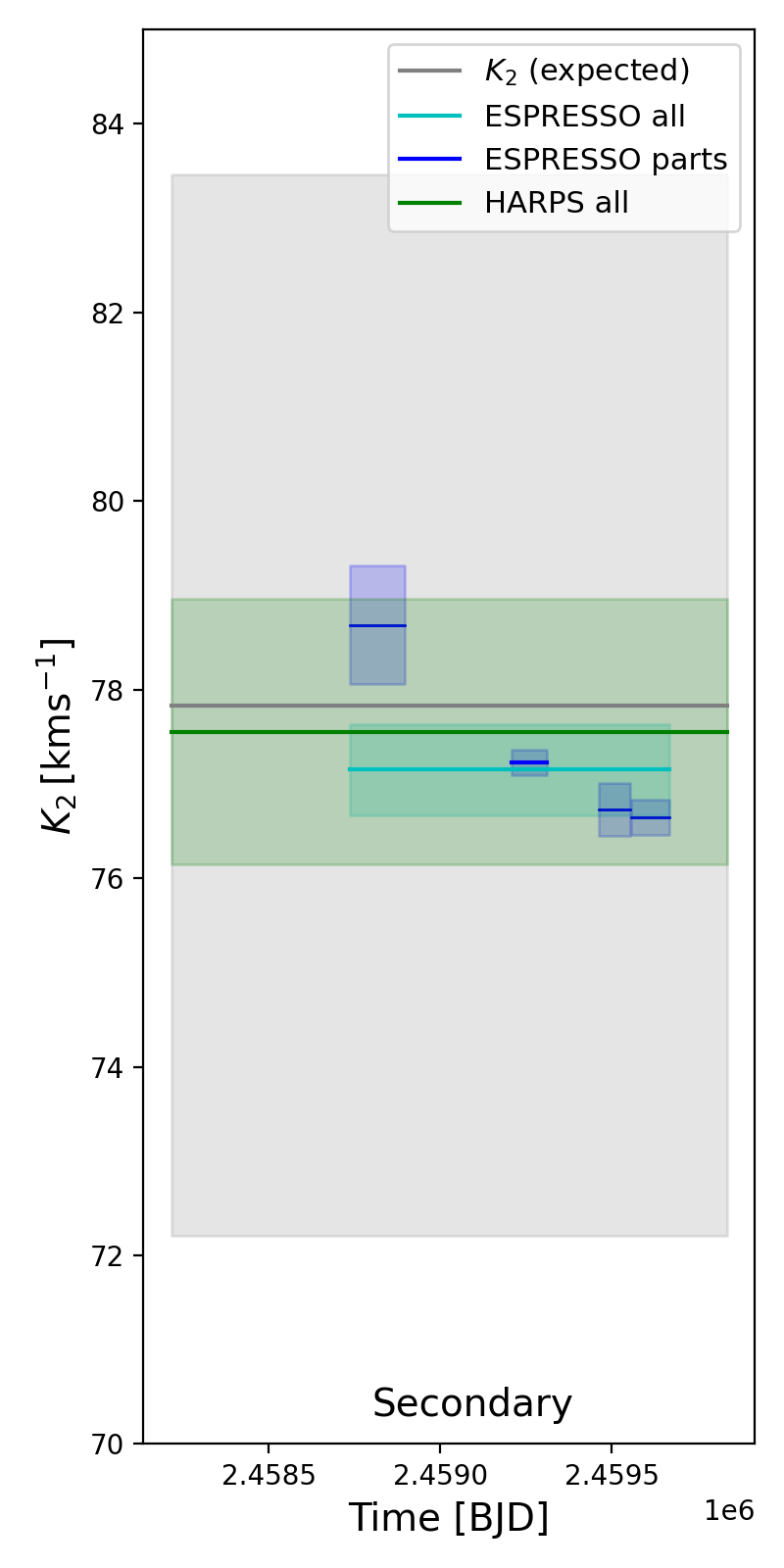}
    \caption{MCMC Fit results for the measured semi-amplitude from partial data samples for both ESPRESSO and HARPS instruments. Left panel: Primary components, Right panel: Secondary component. Uncertainties for each partial sample are fit errors, not taking systematic errors into account. Cyan and Green shaded areas: Uncertainties of all HARPS and ESPRESSO measurements, taking systematic errors into account. The Gray shaded area shows the expected uncertainty for the secondary semi- amplitude.}
    \label{fig:Uncs}
\end{figure}

For each of these parts, we create a CCF map and derive the best-fitting parameters for $K_{2}$ and $V_{\rm rest,2}$ using the MCMC sampler of {\tt Saltire}. Fig.~\ref{fig:Uncs} shows the individual measurements, returned for each partial CCF map. We then estimate the resulting systematic error as the standard error of the four individual measurements. Here we apply Bessel's correction, to derive the standard deviation of this small sample. We note that the standard deviation from $n=4$ measurements has itself a fractional error of $1/\sqrt{2n-2}=41$~per~cent \citep{Topping1972}. In Table~\ref{tab:uncertainties}, we show the combined error for both parameters as the quadratic sum of the total fit error, and the systematic error. For the ESPRESSO data, we find that the systematic uncertainties for the secondary are about seven times larger than the fit errors. For HARPS we expect larger systematic uncertainties due to the low detection significance. We find that dividing the data set in four parts, as described above, results in a detection significance on the order of 1-${\sigma}$ per part, making it impossible to run a meaningful fit. As described above, our measurements from ESPRESSO data show that the fit error is seven time smaller than the systematic uncertainties. Assuming the systematic uncertainties for the HARPS data set are similarly underestimated, we thus, can use the fit error, obtained from the full HARPS sample to place an estimate for the HARPS uncertainties. Increasing the fit error of the full HARPS data set, by a factor of seven leads to an uncertainty estimate of $3\,\rm km\,s^{-1}$ , which we add in quadrature to the fit error.


Fig.~\ref{fig:Uncs} shows the results for the primary component. We find that the absolute deviations to $K_{\rm 1}$, measured by \citetalias{Standing23} are less than $15\,\rm m\,s^{-1}$, with a tentative systematic offset between ESPRESSO and HARPS spectra. deriving the standard error results in 4 $(5.2)\,\rm m\,s^{-1}$ for ESPRESSO (HARPS). Different to the secondary, these estimates of systematic errors are similar or smaller, compared to the fit errors returned by {\tt Saltire}. Taking these errors into account, the HARPS measurement would be statistically identical to $K_{\rm 1}$, while the ESPRESSO measurement would be slightly offset by about 2-${\sigma}$ 

Deriving the standard error requires the four measurements being statistically independent, which is true for white noise dominated data. With a detection significance of >300-${\sigma}$ for the primary, we are in a high-signal-to-noise regime. We, thus, expect aliases from spurious correlations of the M-dwarf spectrum with the line mask (henceforth denoted as ``wiggles'') being the dominating source of these systematics. In this case the measurements are no longer independent, since similar orbital phases are present in each part. Thus, for the primary, we underestimate the uncertainty using the standard error. We discuss the effect of such wiggles in \cite{Sebastian23b}, which limits the accuracy from the fitting method to $20\,\rm m\,s^{-1}$ for noise-less data. We note that the measurements of the four parts of each instrument vary similarly at the $20\,\rm m\,s^{-1}$ level. We thus estimate the systematic uncertainties for $K_{\rm 1}$ and $V_{\rm rest,1}$ for the primary as the absolute error, which we add in quadrature to the fit errors.  Our measurements of $K_{\rm 1}$ for both ESPRESSO and HARPS agree within this accuracy.

For $V_{\rm rest,2}$, we find about four times larger systematic errors for ESPRESSO data, compared to the fit errors. For HARPS, we similarly cannot determine the systematic errors directly. We thus, add a noise term of 1.1${\rm km\,s^{-1}}$, which is four times the fit error of full HARPS data set, to account for systematic noise. For both HARPS and ESPRESSO data, we find that the secondary rest frame velocity is red-shifted of about $1.5\pm0.4\,{\rm km\,s^{-1}}$ compared to the primary ($V_{\rm rest,1}$). Since this is a relative red-shift, compared to the primary's rest-frame, it should be independent from the instrumental offsets. This red-shift is likely a superposition of factors such as aliases from the analysis, like the different line-list, used for the primary and secondary components, as well as from physical effects, such as convective blue-shift and gravitational red-shift. 

\begin{table}
	\centering
	\caption{Uncertainties of retrieved parameters from systematic analysis. Errors in brackets are the precision (first) and the systematic (second) errors.}
	\label{tab:uncertainties}
	\begin{tabular}{lcc}
    \hline
    Instrument & Parameter & error [${\rm km\,s^{-1}}$]
    \\\hline
    ESPRESSO & $K_{\rm 1}$ & 0.0195 (0.0046, 0.019) \\
     & $K_{2}$ & 0.479 (0.078, 0.472) \\
     & $V_{\rm rest,1}$ & 0.0051 (0.00295, 0.0042) \\
     & $V_{\rm rest,2}$ & 0.26 (0.06, 0.25) \\\hline
    HARPS    & $K_{\rm 1}$ & 0.0205 ( 0.0046, 0.020) \\
     & $K_{2}$ & 3.03 (0.430, 3.0$^{*}$) \\
     & $V_{\rm rest,1}$ & 0.0048 (0.0032,0.0036) \\
     & $V_{\rm rest,2}$ & 1.13 (0.269, 1.1$^{*}$)\\
    \hline
    $^{*}$ assumed systematic error
    \end{tabular}
    \\
\end{table}

\subsection{Uncertainties From Orbital Parameters}

Similar to typical HRCCS analyses, we kept the orbital parameters fixed, ignoring their uncertainties. In this section, we investigate the effect of these uncertainties on the measured signal position.
To estimate these uncertainties, we first draw a new set of 425 orbital parameters ($P,T_{\rm 0,peri},e,\omega$) from the uncertainties, reported in \cite{Standing23} assuming a normal distribution for each parameter. To measure the signal position, we then need to derive a CCF map for each of these sets of orbital parameters. To generate these CCF maps in a computational efficient way, we make use of the robustness of {\tt Saltire}, which allows us to derive precise position parameters also for CCF maps with a smaller sampling of $K_{2}$ \citep{Sebastian23b}. Thus, we decide to scan the $K_{2}$-range for only 18 steps with a spacing of $2.5\,{\rm km s^{-1}}$. Additionally, we only use the second partial sample with 25 spectra, introduced in Sec.~\ref{uncs}. It has the highest detection significance for the secondary (SNR = $8.54\pm0.30$), which is perfectly suitable to accurately measure the signal position for different orbital parameters.
We first measure the signal position of this reduced sample for the original orbital parameters using {\tt Saltire}, which is statistically identical to the full map for this sample. We then measure the signal position for all 425 CCF maps and investigate the resulting parameter distribution for $K_{2}$, $V_{\rm rest,2}$, and the CCF contrast ($\Sigma = A_1+A_2$)

\begin{table}
	\centering
	\caption{Propagated systematics from orbital uncertainties on the signal position on a partial sample of ESPRESSO spectra. First: Fit of a CCF map with fixed orbital parameters, Second: Mean and RMS from fitted parameters to CCF maps with different orbital parameters.}
	\label{tab:orbit}
	\begin{tabular}{lc}
    \hline
    
    \multicolumn{2}{c}{Single orbit}  \\
    Parameter & Fit result \\\hline
    
    $K_{2} [{\rm km\,s^{-1}}]$ & $77.30\pm0.11$ \\
     $V_{\rm rest,2} [{\rm km\,s^{-1}}]$ & $31.88\pm0.08$ \\
     $A1+A2 [{\rm ppm}]$ & $247.41\pm0.98$ \\ \hline
     
    \multicolumn{2}{c}{Orbital parameter sample}  \\
    Parameter & Systematic result \\\hline
     $K_{2} [{\rm km\,s^{-1}}]$ & $77.312\pm0.014$ \\
     $V_{\rm rest,2} [{\rm km\,s^{-1}}]$ & $31.89\pm0.01$ \\
     $A1+A2 [{\rm ppm}]$ & $247.45\pm0.47$ \\ 

    \hline
    \end{tabular}
\end{table} 

In Table~\ref{tab:orbit} we present the best-fitting parameters as the 50th percentile, and the resulting uncertainties by averaging the 15.8655/84.1345 percentiles of the distribution. We do not see any significant correlation between the resulting parameters and the orbital parameters, with only a marginal correlation between $K_{2}$ and the orbital parameters $T_{\rm 0,peri}$ and $\omega$ (see Fig.~\ref{fig:A_orbit}). The results show that including the orbital uncertainties will only add systematic uncertainties on the 10~per~cent level of the fit error, returned by {\tt Saltire} and are, thus, negligible for this analysis. 

\section{Effects Of Post-Processing}\label{SVD_effects}

The measured systematic uncertainties could be partially caused by the post-processing of the data. As described in Sec.~\ref{svd}, we use an SVD detrending to remove the stellar lines from the primary star from the spectra.

\subsection{Phase dependent SVD degradation}

During the SVD detrending, all spectra are aligned to the primary's rest-frame. By design, parts of the secondary's signal at orbital phases, which are aligned in the primary's rest-frame, might be removed as well. This effect has been demonstrated for phase-resolved observations of the giant exoplanet KELT-9b \citep{Pino22}. Any such phase-dependent degradation can cause systematic changes in the measurements of the combined data. Furthermore, since the applied SVD detrending basically subtracts parts of the spectra (see Sec.~\ref{svd} for details), any correction for such degradation is very difficult, if the absolute phase-dependent signal strength is unknown. In this section, we investigate the magnitude of this data degradation on our measurements of semi-amplitude and rest velocities.

\begin{figure}
	
        \includegraphics[width=0.95\linewidth]{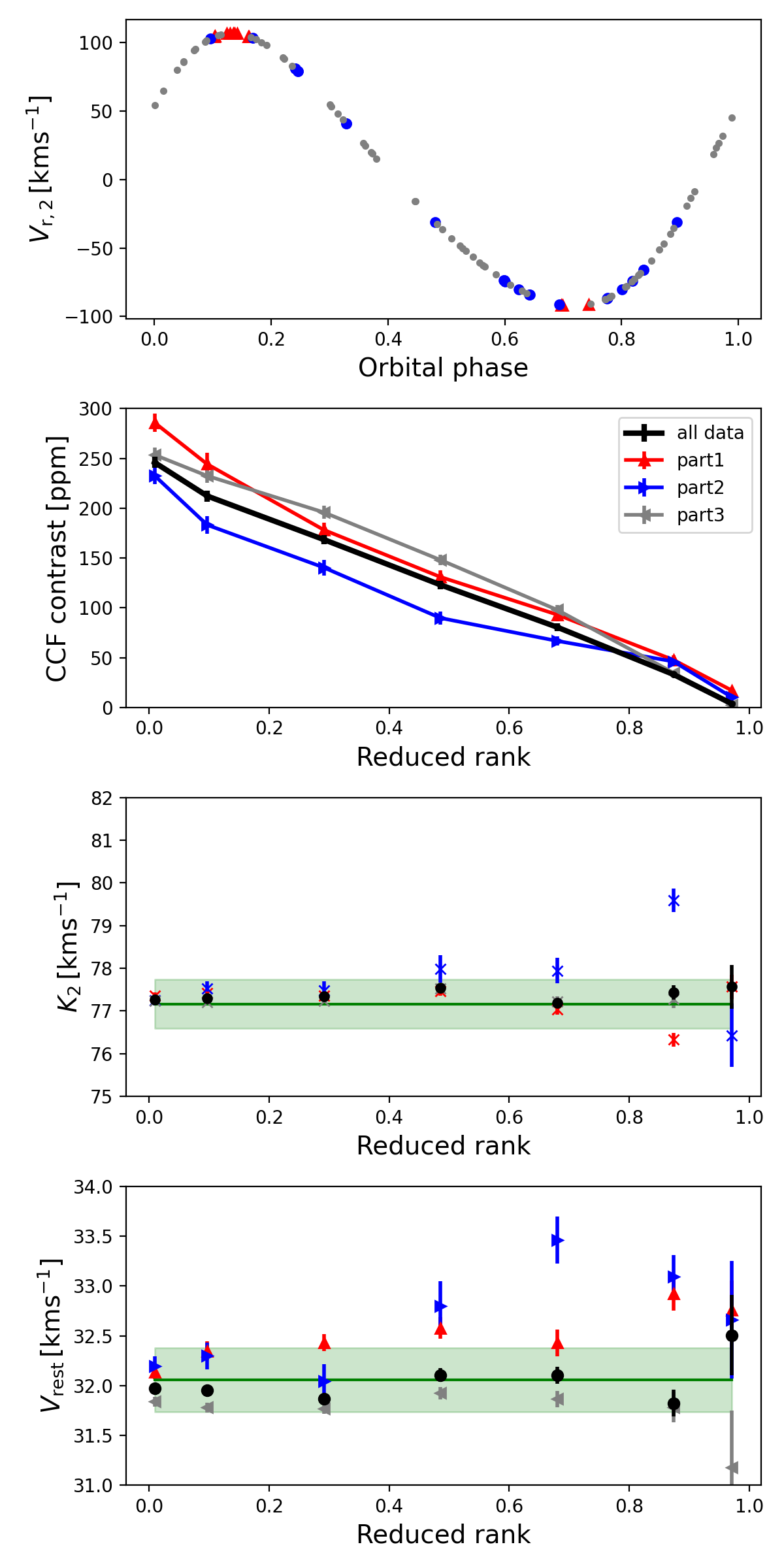}
    \caption{Effects of SVD detrending for ESPRESSO data. Upper panel: Selection of measurements according to their alignment in the primary's rest frame: Red: aligned with several other measurements (part1), Blue: aligned with one other measurement (part2), Grey: not aligned (part3). Similar parts and colour coding are used for all other panels with black points indicating the full data set. Second panel from the top: Measured CCF contrast versus reduced rank. The linear trend indicates a white noise-like degradation without measurable phase-dependence. Two bottom panels: Measured semi-amplitude and rest-velocity of the secondary, versus reduced rank. Green shaded area: Uncertainties including systematic errors. Measurements for all parts agree within the uncertainties for small reduced ranks. }
    \label{fig:rank_test}
\end{figure}

Fig.~\ref{fig:rank_test} upper panel shows the expected position of the secondary in the primary's rest frame. To estimate any differential effect of the SVD detrending for observations at different phases, we split them into three parts, which we expect to have a different degradation profile. Mostly affected by the degradation should be spectra of the secondary, which are aligned in the primary's rest frame, hence are expected to share a similar relative velocity. To quantify this, we define spectra to be aligned, if their relative velocities do not exceed the pixel resolution of ESPRESSO data ($0.5\,{\rm km \,s^{-1}}$). For each spectrum, we first count how many other spectra are within this range. We then select ({\it i}) well aligned spectra which share a similar velocity with two or more other spectra (part1), ({\it ii}) spectra which are weakly aligned that share a similar velocity with only one other spectrum (part2), and ({\it iii}) spectra which do not have other spectra within this range and are, thus, expected to be not aligned in the primary's rest frame (part3). This results in three parts with 13, 23, and 67 spectra respectively. In Fig.~\ref{fig:rank_test}, the spectra with largest expected alignment in the primary's rest frame (part1) appear at phases when the radial velocity of the secondary is changing slowly. 

The phase-dependent degradation might also depend on the SVD detrending, hence the number of excluded components ($k$). We test different SVD detrendings using a fixed number of $k$ for the whole data set instead of the automatic selection, used to clean the data described above. The input matrix $A$ has a rank $R$ of 103, which is the number of spectra used. To scan the full range of possibly excluded SVD components, we explore seven different values of $k$ ($k=$ 1, 10, 30, 50, 70, 90, and 100) to detrend the data. To make this analysis more portable to different data sets as well as to exoplanet studies, we remove the dependence from the absolute number of spectra by defining the reduced rank as $k/R$. This will allow us to compare our results with other works, which use different matrix dimensions for $A$.

We now repeat the SVD detrending for the whole 103 spectra to create seven post-processed datasets, using the seven defined values for the reduced rank ($k/R$). We then create 28 (7x4) individual CCF maps. Each for the seven differently post-processed data, using (i) the whole data set, and (ii) each of the three partial samples of differently aligned spectra. To each of these CCF maps we then fit the {\tt Saltire} model to derive the best-fitting model parameters.

The second panel of Fig.~\ref{fig:rank_test} shows, that the measured CCF contrast of the secondary is decreasing with increasing reduced rank due to the SVD degradation. This decrease in signal strength follows largely a linear trend, which is consistent with an uncorrelated white noise array, which would result in a homogeneous decrease of the secondary's signal at each orbital position. This is supported by the observation that even the most aligned spectra (part1) are similarly degraded like the non aligned spectra (part3). We observe a CCF contrast difference at the 20~per~cent level for small $k$. A variation at this level is expected from correlated noise (see Section~\ref{model_comp}). CCF maps from each partial sample contain different realisations of white noise, causing such differences. Thus, we can not exclude minor, phase-dependent degradation below this noise level at this point.

The two lower panels of Fig.~\ref{fig:rank_test} show that CCF maps of the whole data set - independent from the reduced rank - return consistent parameters for the semi-amplitude and rest velocities (within the uncertainties derived in Sec.~\ref{uncs}). Furthermore, we find consistent parameters for the three partial samples for a wide range of reduced ranks of 0.3 and below. Clearly the partial samples show a larger scatter, which is simply caused by the smaller number of spectra used for part1 and part2, resulting in a higher noise content. Therefore,  we do not find for our data any correlation of the measured CCF signal position with the selected SVD rank.


\subsubsection{Comparison to simulated data} \label{model_comp}

We did not find evidence for phase-dependent degradation for our ESPRESSO data. Nevertheless, such effects might be covered by the noise of the data and be of importance for high SNR observations. In this section, we compare our data with simulated observations with similar and different noise budget.

We derive seven sets of simulated observations of TOI-1338/BEBOP-1. The primary and M-dwarf secondary are simulated with PHOENIX model spectra \citep{Husser13}. For the primary, we use $T_{\rm eff}=6000\,{\rm K}$, $\log\,g_\star=4.0$ and for the secondary with $T_{\rm eff}=3300\,{\rm K}$ and $\log\,g_\star=5.0$. For both was assumed a solar metallicity of ($\rm [Fe/H]=0.0$). We first correct the model wavelengths from vacuum to air, following \cite{Morton00}, then we match the spectral resolution to ESPRESSO ($R\sim140\,000$) using the implementation in the {\tt iSPec} package \citep{Blanco-Cuaresma14}, which convolves the spectra with a Gaussian kernel. In a next step, we multiply the flux calibrated model spectra by the squared radius known from eclipse light curve modelling to derive the residual flux of the secondary in our EBLM model. Nevertheless, we noticed, that the resulting CCF contrast of the modelled secondary would be 35\,per~cent smaller, compared to the contrast we measure from ESPRESSO data. Such a difference can be caused by several effects, such as (i) an insufficient flux calibration of the used model spectra, (ii) a mismatch in the used line mask with the spectral lines of the model, and (iii) slightly different stellar parameters of both stellar components. 

To compare the SVD degradation, the simulated flux and CCF contrast should optimally match to our data. Therefore, in a next step, we normalise the continuum of the model spectra by dividing them though a polynomial function. We then simulate the continuum of each component using a simple blackbody approximation at the modelled effective temperature. This correction decreases the average flux of the primary by about 14\,per~cent and increases that of the secondary by about 30\,per~cent, resulting in an agreement between the modelled and observed CCF contrast better than 10\,per~cent. This is which is what we aim for, to obtain comparable results for the SVD degradation. We also try to replace the blackbody estimation with SED spectra derived by \cite{Coelho14}. Using similar stellar parameters for the primary SED, the average flux decreases by 9\,per~cent. Using an SED for the secondary with an effective temperature of $T_{\rm eff}=3400\,{\rm K}$ results in an average increase of the secondary's flux by 24\,per~cent. Both corrections lead to an modelled CCF contrast, which is underestimated by about 15\,per~cent, compared to the data. We thus use for our EBLM model the simple blackbody approximation as it matches the observed results best. In the case that the CCF contrast difference is dominated by a line mask mismatch, rather than the SED, our correction would thus lead to an overestimation of the secondary's flux by about 30\,per~cent. In other words, we would minimally overestimate (by about 15\,per~cent) the average SNR (SNR = 0.09) for the secondary per resolution element in each ESPRESSO spectrum. The analyses, presented in this and the following sections are therefore unaffected by our choice of continuum normalisation.

Each spectrum is then shifted by linear interpolation to match the secondary's reflex motion using the primary's stellar and orbit parameters (see Table~\ref{tab:bin_par}). We insert $K_{\rm 2,model}= 77.57\,{\rm km\,s^{-1}}$ which is a random value close to the measured semi amplitude. The first simulation (1) takes the advantage of excluding the primary component from the model and, thus, is a noise-less spectrum of the M-dwarf component only. This completely avoids the necessity of SVD detrending. 

For our second simulation (2), we use the full model with both components. To derive a noise model for ESPRESSO, we first use one observed ESPRESSO spectrum normalised at $550.0\,{\rm nm}$. Then, we smooth it using a sliding window of 600 pixels \footnote{adopted from the \hyperlink{https://github.com/kevin218/Eureka/blob/main/src/eureka/lib/smooth.py}{smooth} function of the Eureka pipeline as of the 2022-07-25} which has been sigma clipped with a threshold of $10\,{\rm \sigma}$ and 5 iterations. This noise model can now be used to create observations at different SNR, by multiplying it with white noise corresponding to each single observation. We add white noise to each spectrum, corresponding to the observed SNR, of our real ESPRESSO observations, resulting in a similar average SNR. 

The SNR of the ESPRESSO observations follow roughly an uniform distribution due to changing atmospheric conditions. We thus simulate observations at higher SNR (Simulations 3 - 6), by assuming uniform distributions for the SNR, as listed in Table~\ref{tab:SVD_deg}. Not all higher SNR would be possible for real observations of TOI-1338/BEBOP-1 due to detector saturation. Nevertheless, we include them to cover a wider range of SNR of the secondary, which is interesting for binaries with lower contrast ratios. Finally, we simulate a noise-less observation (Simulation 7) of the full model. For simulations 2-7, we repeat the steps from the previous section by applying the SVD detrending with different $k$ and splitting the data in three parts of different correlation. We finally measure the residual signal of the secondary for each resulting CCF map. 

For simulation\,2, the results show a very similar decrease of CCF contrast with reduced rank, compared to the observed data (see Fig.~\Ref{fig:A_erosion}). Furthermore, we  see a similar scatter for the measured semi-amplitude and rest velocity, as well as similar CCF contrast differences at the 20~per~cent level for small $k$. This scatter decreases for higher SNR in our simulations\,3 and 4.

These three simulations (2 - 4) are all well in agreement to a white noise dominated degradation, without any significant changes between the different partial samples. Only at very high SNR in our simulations\,5 to 6, do we observe significant and nonlinear degradation with the reduced rank, meaning the correlated signal is being detrended by the SVD also for larger $k$. The explanation for this behaviour can be found in the secondary's signal, which is with an average SNR>1 not anymore buried in the noise of the simulated observations. We now see a clear trend for highly aligned spectra (part1) being more degraded than less aligned spectra (part3), which is different to our ESPRESSO data and the simulations for lower SNR. In principle, we observe now the expected phase-dependent degradation of the secondary's signal with reduced rank.  

Our simulations, thus, confirm our previous assumption that the secondary's signal is in the ESPRESSO data only negligibly affected by phase-dependent signal degradation from SVD detrending, but is degraded homogeneously for larger reduced ranks. We make use of this aspect in Sec.~\ref{phasecurve}.

\begin{table*}
	\centering
	\caption{Dependence of the measured CCF contrast from the number of excluded SVD components - measured as reduced rank ($k/R$) - for simulated observations. The signal amplitude is the fractional CCF contrast, compared to a non-detrended simulated observation. Primary and secondary SNR for each simulation are given per resolution element of individual observations. The reduced rank of 0.1 would correspond to a $10^{\rm th}$ of the number of observations (e.g. an $k$ of 3 for 30 spectra). Simulation 7 assumes noiseless data, thus showing the special case of maximal degradation possible for this data-set.}
	\label{tab:SVD_deg}
	\begin{tabular}{lcccccc}
    \hline
    
    Simulation & \multicolumn{2}{c}{Primary SNR} & Secondary SNR & \multicolumn{3}{c}{Signal amplitude [\%]} \\
    & SNR range & average & average & red. rank: 0.01 & 'auto' rank &red. rank: 0.1 \\\hline
    2 & [15,58] & 40 & 0.09 & $99.6\pm2.6$ & $99.3\pm2.7$ & $95.4\pm2.5$\\
    3 & [40,140] & 96 & 0.2 & $96.9\pm1.2$& $94.2\pm1.2$ & $89.4\pm1.1$\\
    4 & [80,247] & 175 & 0.4 & $94.1\pm0.8$& $91.2\pm0.8$ & $85.6\pm0.8$\\
    5 & [401,568] & 490 & 1.0 & $92.3\pm0.6$ & $83.1\pm0.7$ & $61.1\pm0.6$\\
    6 & [4001,4178] & 4084 & 8.8 & $77.9\pm0.5$ & $66.6\pm0.5$ & $44.9\pm0.8$\\
    7 & -- & $\infty$ & $\infty$ & $76.1\pm0.5$ & $23.6\pm0.5$ &  $31.1\pm0.4$ \\

    \hline
    \end{tabular}
\end{table*}

\subsection{Signal degradation at low SVD ranks} \label{corr_noise}

We established, in the previous section, that the signal degradation from the SVD detrending is a function of the secondary's SNR in each single observation. Thanks to the simulated observations from the previous section, we can derive in what amount the secondary signal is actually degraded in our measurements. Our simulation 1 is not affected by the SVD detrending, thus can be used as direct reference.

\begin{figure}
	
        \includegraphics[width=\linewidth]{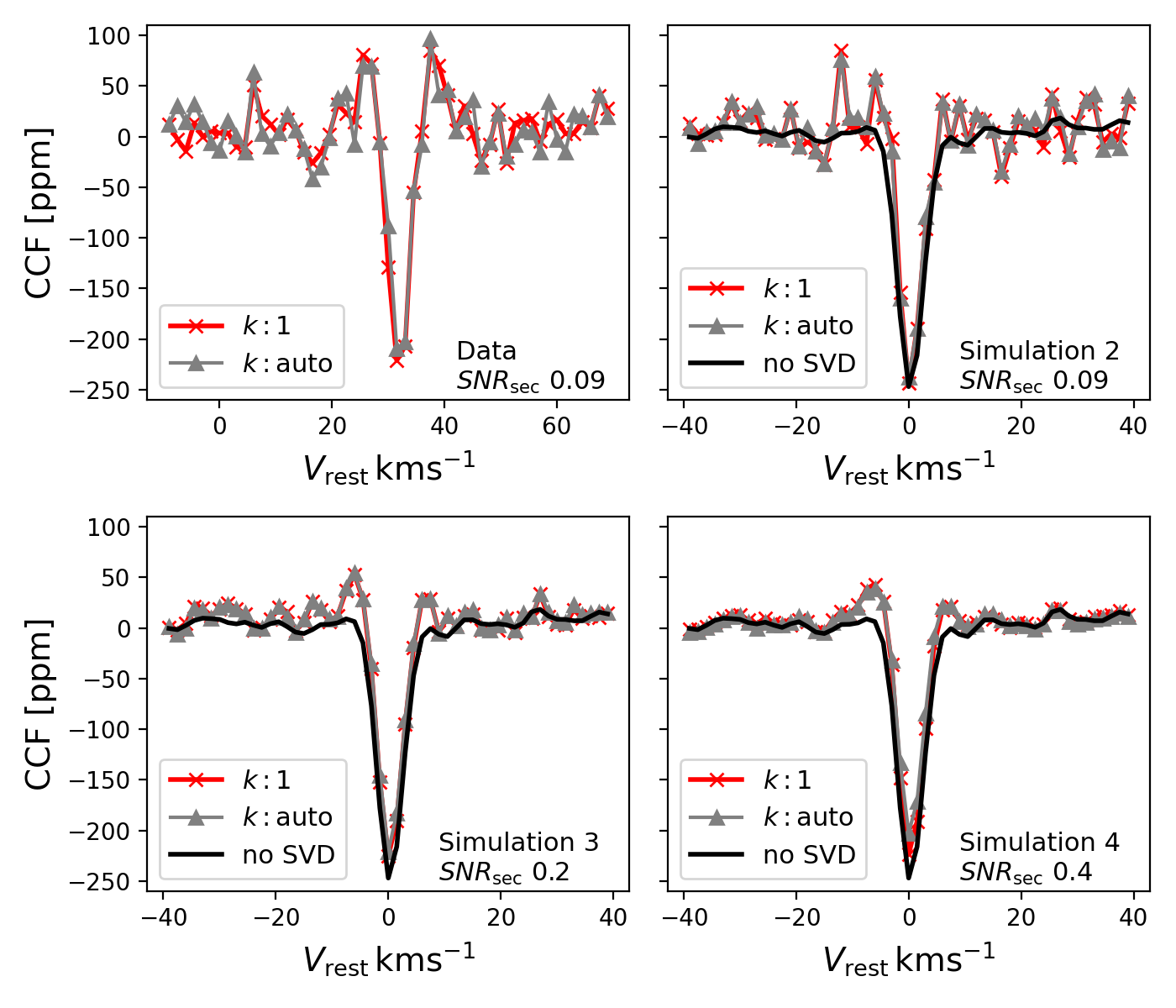}
    \caption{Slice through CCF map at semi-amplitude of maximum CCF contrast for different SVD detrending. Upper panel: Left: combined ESPRESSO data. Red: first SVD component removed, grey: using the effective rank to exclude SVD components, The SNR of the secondary in each individual observation is indicated. Right: Simulation of combined ESPRESSO data with similar SNR and SVD detrending. The Black line shows Simulation 1 without SVD detrending for comparison. Lower panels: Similar to above for simulations 3 and 4 with larger SNR of the secondary. The comparison to simulation 1, allows us to estimate the degradation as a function of the secondary SNR and excluded SVD components.}
    \label{fig:profile}
\end{figure}

Fig.~\ref{fig:profile} shows the CCF profile of the secondary at maximum contrast for the ESPRESSO data, as well as for the simulated observations for different SVD detrending. We show the profile $k=1$ (only the first components excluded) as well as for the `auto' detrending, based on the effective rank (See Sec.~\ref{svd} for details). Since we showed that the secondary's signal is strongly degraded as soon as its SNR is on the order of one or larger, we only show the profiles for simulations where the secondary's SNR is much smaller than one.

The resulting comparison is listed in Table~\ref{tab:SVD_deg}. We find that the measured CCF contrast for our second simulation is $99.6\pm2.6$~per~cent after removing the first SVD component only. The detrending, using the 'auto' rank results in $99.3\pm2.7$~per~cent. Both are identical to the unprocessed CCF contrast, within the measured precision. From this we can conclude that the measured CCF contrast for our ESPRESSO data is unlikely to be degraded more than the uncertainties of the measurement.

Studies of exoplanet spectra typically use fewer spectra than we have used here, and suffer from imperfect alignment of the telluric lines. This results in the necessity to apply SVD with more components being removed. A typical value for $k$ corresponds to about 10~per~cent of the number of spectra taken. In Table~\ref{tab:SVD_deg}, we show the simulated degradation for the reduced rank of 0.1, which corresponds to $k = 10$ for our simulation. If only 50 spectra were taken, this would correspond to $k = 5$. We show that in this case 5 - 10~per~cent of the CCF contrast will be degraded, even if the secondary's SNR is $\le 0.2$.

Both profiles, of the ESPRESSO data as well as of our second simulation (2), show noise structures of very similar amplitude and spacing. These are expected, since we assume a similar SNR and use a noise model derived from the observed data.

These noise structures are very similar despite different SVD detrending. This is not surprising as the noise structure originates from white noise of the individual spectra, which becomes correlated due to the K-focusing process. The SVD detrending does remove data which are aligned in the primary's rest frame and has only a minimal effect on the non-aligned white noise. Thus, the white noise from different SVD detrendings become similarly correlated noise structures. 

For our simulations with higher SNR, these structures are shallower, which is expected from white noise. Our simulations suggest that with an increment of SNR of the secondary, the systematic errors are less dominated by correlated white noise. If the SNR of the secondary is larger than 0.3, we enter a high SNR regime. As shown in Fig.~\ref{fig:profile} for our simulation 4, in this high SNR regime the systematics follow the structures of the (black) undetrended profile. This means measurements for such high SNR secondaries will not only be dominated by white noise, but also by spurious correlations of the secondary spectrum with the line mask, so-called wiggles, as we discuss in \cite{Sebastian23b}. The measured semi-amplitude $K_{2}$ for this simulation agrees within $7\,{\rm m\,s^{-1}}$ to the injected semi-amplitude, which is in agreement with the measured uncertainty of $39\,{\rm m\,s^{-1}}$.

Fig.~\ref{fig:profile} also shows the expected side-lobes from the M-dwarf CCF in the ESPRESSO data. These are not apparent in the unprocessed simulated observation (see simulations, black line). Despite such side-lobes are not part of our simulation, we note that for our higher SNR simulations, some residual structures are still persisting close to the main CCF signal, forming side-lobe-like structures. These structures have to be residual structures from the post-processing. Thanks to the high SNR of the secondary in our simulation 6, these structures become visible in individual CCFs. In Fig.~\ref{fig:A_side_lobes}, we show these structures for simulation 6 in the rest frame of the primary. The artificial side-lobes are clearly aligned with the primary's rest frame, making it apparent to be a residual from the signal degradation of the secondary from the SVD detrending.

\section{Dynamical masses}\label{masses}

We can use the measurements for the secondary's semi amplitude, to derive the mass ratio ($q = K_1/K_{2}$) between both stars. Since the inclination ($i$) is precisely known from light curve analysis \citep{kostov20}, we can convert these to model-independent dynamical masses of both the primary and the secondary star.

We follow the IAU recommended equations \citep[Table 3 in][]{2016AJ....152...41P} to derive the dynamical masses, and semi major axis directly from the measured parameters as listed in Table~\ref{tab:bin_par}. We note that the differences for $K_1$ of few $\rm m\,s^{-1}$ to the more precise measurements from \citetalias{Standing23} do not affect the resulting masses and uncertainties. We, thus, use $K_1$, measured in this work to derive the stellar masses. We also compare the dynamical masses to literature values, as listed in Table~\ref{tab:bin_par}. Our measurements agree within 1-$\sigma$ of the uncertainties presented in those previous works. Furthermore, we achieve about four times smaller uncertainties for the stellar masses of both components.

As discussed before, the HARPS detection is dominated by white noise, which increases the systematic uncertainties. Nevertheless, we find that the Masses, derived from HARPS measurements ($M_{1}=1.11\pm0.11\,M_{\rm \sun}, M_{2}=0.310\pm0.019\,M_{\rm \sun}$) statistically agree with the results obtained with ESPRESSO.

\begin{table}
	\centering
	\caption{TOI-1338/BEBOP-1 - Binary parameters and dynamical masses. Comparison between radial velocity measurements as single lined binary (SB1) and double lined binary (SB2, this work).}
	\label{tab:bin_par}
	\begin{tabular}{ll}
    \hline
    parameter & value\\\hline
    $V_{\rm mag}$ & $11.72\pm0.02^{\dagger}$ \\
    $T_{\rm mag}$ & $11.45\pm0.02^{\dagger}$ \\
    $T_{\rm eff,1} [{\rm K}]$ & $5990\pm110^{\dagger}$ \\
    $\log g_1 [{\rm cgs}]$  & $4.0\pm0.08^{\dagger}$ \\
    $\rm[Fe/H]$ & $0.01\pm0.05^{\dagger}$ \\
    $T_{\rm eff,2} [{\rm K}]$ & $3317\pm67^{\dagger}$ \\
    $R_1 [{\rm R_{\sun}}]$ &  $1.299\pm0.025^{\dagger}$ \\
    $R_2 [{\rm R_{\sun}}]$ &  $0.3015\pm0.0058^{\dagger}$ \\
    $P [{\rm d}]$ & $14.6085579\pm0.0000057^{\ast}$ \\
    $T_{\rm 0,peri} [{\rm BJD}]$ & $2\,458\,206.16755\pm0.00071^{\ast}$ \\
    $e$ & $0.155522\pm0.000029^{\ast}$ \\
    $\omega [{\rm rad}]$ & $2.05549\pm0.00030^{\ast}$ \\
    $i [{\rm ^{\circ}}]$ & $89.658\pm0.146^{\dagger}$\\\hline
    \multicolumn{2}{c}{SB1 measurements from literature}  \\
    parameter & value\\\hline
    $K_{1} [{\rm km\,s^{-1}}]$ & $21.61764\pm0.00073^{\ast}$\\
    $a [{\rm AU}]$ & $0.1321\pm0.0025^{\ast}$\\
    $M_{1} [{\rm M_{\sun}}]$ & $1.038\pm0.069^{\dagger}$\\
    $M_{2} [{\rm M_{\sun}}]$ & $0.2974\pm0.0116^{\dagger}$ \\
    $M_{2} [{\rm M_{\sun}}]$ & $0.313\pm0.012^{\ast}$ \\\hline

    \multicolumn{2}{c}{SB2 measurements using ESPRESSO}  \\
    parameter & value\\\hline
    $K_{1} [{\rm km\,s^{-1}}]$ & $21.6037\pm0.0195$ \\
    $K_{2} [{\rm km\,s^{-1}}]$ & $77.16\pm0.479$\\
    $a [{\rm AU}]$ & $0.1310\pm0.0006$\\
    $M_{1} [{\rm M_{\sun}}]$ & $1.098\pm0.017$\\
    $M_{2} [{\rm M_{\sun}}]$ & $0.307\pm0.003$ \\
    

    \hline
    $^{\ast}$ From \cite{Standing23}\\
    \multicolumn{2}{l}{$^{\dagger}$ `Best' parameters from \cite{kostov20}}  \\
    
    \end{tabular}
    \\
\end{table} 

\section{Phase curve of the M-dwarf} \label{phasecurve}

We know that the measured CCF is basically an average of the M-dwarfs lines at the positions of the line mask. Since the M-dwarfs spectrum does not vary with orbital phase and we keep the line mask constant, the CCF contrast is basically a measure of the secondary contribution to the total light. Thus, a constant scaling factor can be used to convert the CCF contrast to the relative intensity 
of the secondary spectrum, averaged over the wavelength range of ESPRESSO.

Additionally, we can make use of two key aspects from the previous analyses. First, the SVD detrending does not introduce a phase-dependent degradation of the CCF signal. Second, the homogeneous degradation due to the SVD detrending can safely be assumed to be less than 1~per~cent of the original signal. The measured CCF contrast from individual measurements can, thus, be treated as an largely unaffected measurement of the secondary's phase curve.

To measure this phase curve, we first post-process all ESPRESSO spectra using an SVD detrending with $k=1$. In a second step, we move the spectra into the secondary's rest-frame, using the measured $K_{2}$ and cross-correlate them with the same M2-dwarf line mask, used in Sec.~\ref{detect}. Instead of combining the resulting CCFs for individual nights, we only combine the CCFs of spectral chunks of each individual spectrum and measure the CCF contrast using a Gaussian function. 

\begin{figure*}
	
        \includegraphics[width=\linewidth]{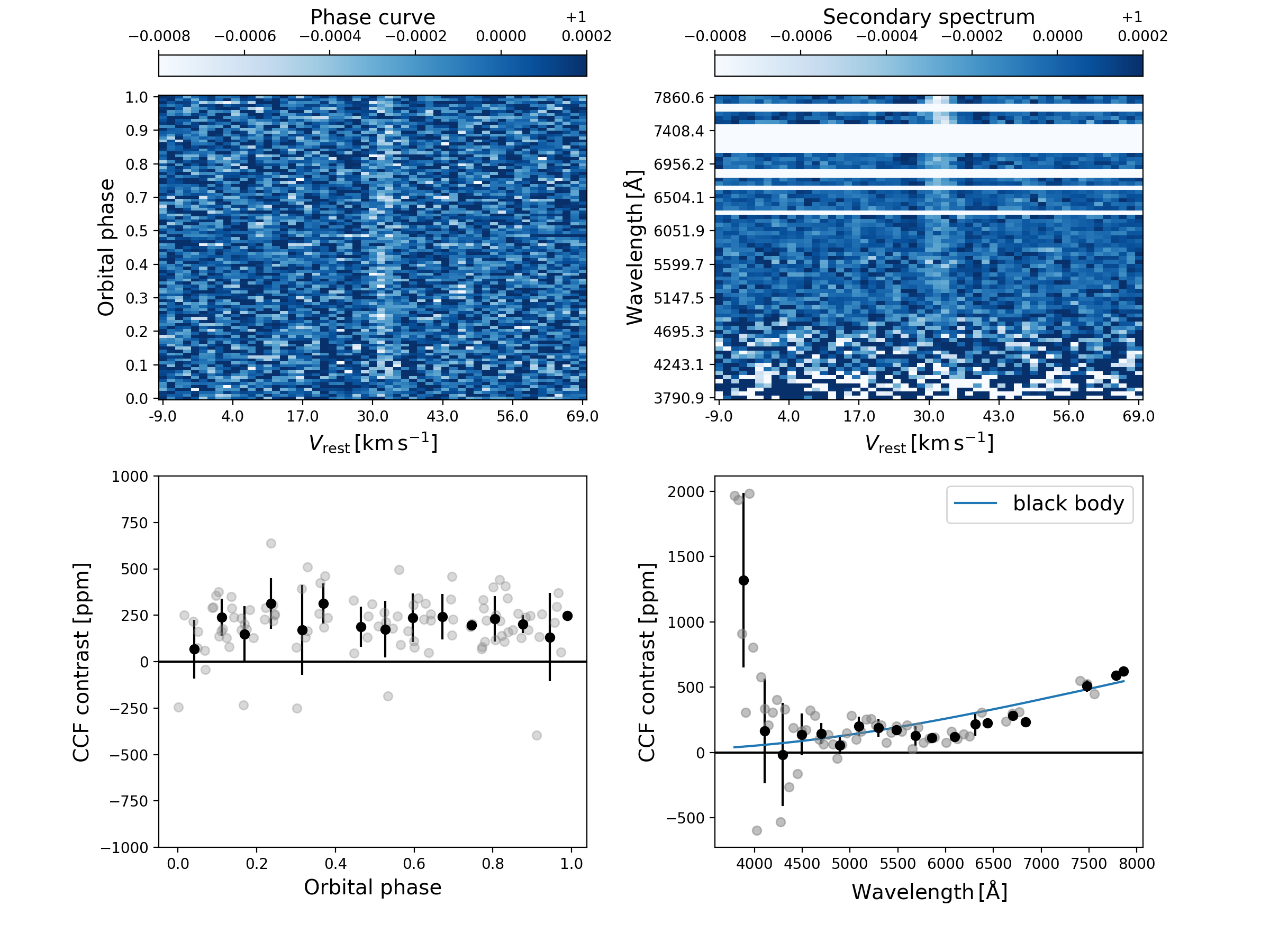}
    \caption{Upper panels, CCF functions in the secondary's rest-frame. Left: as a function of the orbital phase, Right: as a function of the Wavelength. Lower panels, measurements of individual CCF contrasts (grey dots) and binned data (black dots). Left: Phase curve of the M-dwarf secondary, Right: residual flux as a function of wavelength, showing a low resolution spectrum of the M-dwarf. Blue line: residual flux of the secondary, assuming blackbody spectra from parameters listed in Table~\ref{tab:bin_par}, scaled to the average CCF contrast (0.1).}
    \label{fig:phase curve}
\end{figure*}

Fig.~\ref{fig:phase curve} shows the resulting phase curve of M-dwarf secondary. Within the uncertainties, we do not see significant variations of the M-dwarfs contribution as a function of the orbital phase. This confirms the assumption of negligible reflected light and ellipsoidal contributions, used to measure the semi-amplitude with the {\tt Saltire} model.

We finally make use of the splitting in spectral chunks. Instead of combining all CCFs, we measure the combined CCF of each spectral chunk, which reveals the wavelength dependent CCF signal, thus, a low resolution residual spectrum of the M-dwarf. Fig.~\ref{fig:phase curve} shows this spectrum as a function of the average wavelength in each spectral chunk. The large scatter at wavelengths below 450\,nm is caused by the low SNR of the combined spectra. The CCF contrast increases with wavelength, as expected for an M-dwarf, orbiting the G-type primary. To show this, we plot the residual blackbody spectrum, used in Sec.~\ref{model_comp} over the measurements using a scaling factor of 0.1.

\section{Discussion}

We present the measurement of dynamical masses for the planet-hosting high-contrast, eclipsing binary TOI-1338/BEBOP-1. The SNR of the secondary component in each individual spectrum is about 0.09. This low contribution does not allow a direct measurement of the faint secondary signal as a double lined binary. To achieve this, we use high-resolution ESPRESSO and HARPS spectra to which we apply high-resolution cross-correlation methods, typically used to detect atomic or molecular species in exoplanet atmospheres. Similar to our case, such signals are deeply buried in the noise, thus these methods involve two basic steps to detect them: First, the data are detrended by removing the main contributor in the spectra, such as telluric or spectral lines from the primary star. Second, the data are cross correlated in the rest frame of the companion, assuming a range of semi-amplitudes. We call this process K-focusing, since the CCF signal is amplified, when combining all CCFs using the actual semi-amplitude ($K$) of the companion.

Since the main contribution in the optical spectra of TOI-1338/BEBOP-1 originates from the primary star, we detrend spectra in the primary rest frame, using a singular value decomposition (SVD). We show that the uncorrected telluric contribution can be avoided by excluding telluric orders from the analysis, thanks to the wide spectral range of HARPS and ESPRESSO.

Using these methods, we successfully detect the CCF signal of the M-dwarf in both the ESPRESSO and the HARPS data sets with a significance of 11 and $3\,{\rm \sigma}$ respectively. 

We make use of the {\tt Saltire} model \citep{Sebastian23b} to measure the precise semi-amplitude of the M-dwarf secondary, and that of the primary star. This model applies a low parameter fit to CCF signals, amplified by the K-focusing process, and allows us to measure the semi-amplitude taking the actual shape of the signal into account.

We investigate systematic uncertainties from this process. Correlated white noise is typical for the K-focusing process \citep{Hoeijmakers2020,Sebastian23b}. We use the {\tt Saltire} model as a robust diagnostic tool and show that correlated noise is likely the main contribution to the systematic uncertainties, which are on the order of 10 times larger than the precision from the {\tt Saltire} fit of the CCF signal alone. Thanks to the well known binary parameters from the primary star, we show that uncertainties from the orbital parameters are negligible, and only contribute at the 10~per~cent level of the fit errors. 

Taking these uncertainties into account, we derive the dynamical masses for both stellar components with an uncertainty of 1.6~per~cent for the primary and 1~per~cent for the secondary component. The masses are statistically consistent to the published masses by \cite{kostov20,Standing23}. Furthermore, we can securely improve the uncertainties by a factor of four. This shows the importance of this method to derive precise model-independent masses of eclipsing binaries and circumbinary host stars, as our findings are an important confirmation of the stellar models used in their work. Furthermore, measuring dynamical masses will allow to directly measure model-independent masses and radii of any transiting circumbinary planet in the system.

We analyse the effect of the SVD detrending to our measurements, and do not find any systematics in excess to the reported uncertainties from correlated noise. 

By the design of the SVD detrending, we expected a possible phase-dependent degradation of the M-dwarfs signal, but do not find any evidence of this effect in the ESPRESSO data. To further analyse this result, we carry out simulations to analyse the effect of SVD degradation as a function of SNR of the secondary in each individual spectrum. We find that at low SNR, similarly to the ESPRESSO data of TOI-1338/BEBOP-1, the SVD degradation is a homogeneous effect, without measurable phase dependence. From these simulations, we further show that phase-dependent degradation does indeed occur, but only for a SNR of the secondary in the order of one and higher. Such high SNR cannot be reached for real observations of TOI-1338/BEBOP-1 due to saturation of the primary signal. For low contrast binaries, the secondary will reach such high SNR, which most likely means that both components can be measured directly without the need for SVD detrending. 

We show from simulations of similar noise budgets as the ESPRESSO data, that more than 96.6~per~cent of the CCF signal (including the 1-$\sigma$ uncertainty) is likely conserved after the automatic SVD detrending, using the effective rank. Furthermore, we show that for typical exoplanet observations, the CCF signal is likely degraded by 5-10~per~cent. A similar level of degradation has also been reported in the literature \citep[e.g.][]{Cheverall23}. From our findings, we expect that, as long as the SNR of the secondary in each individual observation is less than 0.4, such degradations are likely to be homogeneous, and thus, phase-independent, allowing to trace the secondary's signal over the orbital phase.

We use this fact to derive the phase curve of the M-dwarf secondary of TOI-1338/BEBOP-1, as a function of the CCF contrast versus orbital phase. Despite the non-detection of phase-dependent flux variations, we show the power of this method to derive phase curves of M-dwarf from high-resolution spectroscopy. 

In future studies, we will be able to use these findings to survey EBLM binaries and to measure dynamical masses of late-type M-dwarfs. The recently commissioned NIR spectrograph NIRPS \citep{Bouchy19} can operate in parallel to HARPS allowing simultaneous capture of  spectra of EBLM binaries in both instruments. Since the residual contrast of the M-dwarfs increases for NIR wavelengths, the application of the presented approach to NIRPS spectra will allow the measurement of dynamical masses for a large part of the known EBLM binaries. Furthermore, we note that this approach can be robustly applied to many high-contrast binaries (e.g. a Solar-type secondary to red giant) to measure dynamical masses and spectroscopy phase curves.

\section*{Acknowledgements}

The authors thank the anonymous referee for their helpful comments that improved the quality of the manuscript. This research is also supported work funded from the European Research Council (ERC) the European Union’s Horizon 2020 research and innovation programme (grant agreement n◦803193/BEBOP). MB acknowledges partial support from the STFC research grant ST/T000406/1.
PM acknowledges support from STFC research grant number ST/M001040/1. 
MRS acknowledges support from the UK Science and Technology Facilities Council (ST/T000295/1), and the European Space Agency as an ESA Research Fellow. This Article is based on observations collected at the European Southern Observatory under ESO programmes 103.2024, 106.216B, 1101.C-0721 and 106.212H. This research has made use of the services of the ESO Science Archive Facility.
\section*{Data Availability}

Most data underlying this article is available online as indicated in the specific section or reference. Data obtained with ESO telescopes are available in the ESO Science Archive Facility, at \url{http://archive.eso.org/cms.html}. Corrected data or CCF-maps, underlying this article will be shared on reasonable request to the corresponding author.



\bibliographystyle{mnras}
\bibliography{library} 



\appendix

\section{SVD detrending}
\subsection{Auto SVD detrending of HARPS data}

\begin{figure*}
	\includegraphics[width=0.5\linewidth]{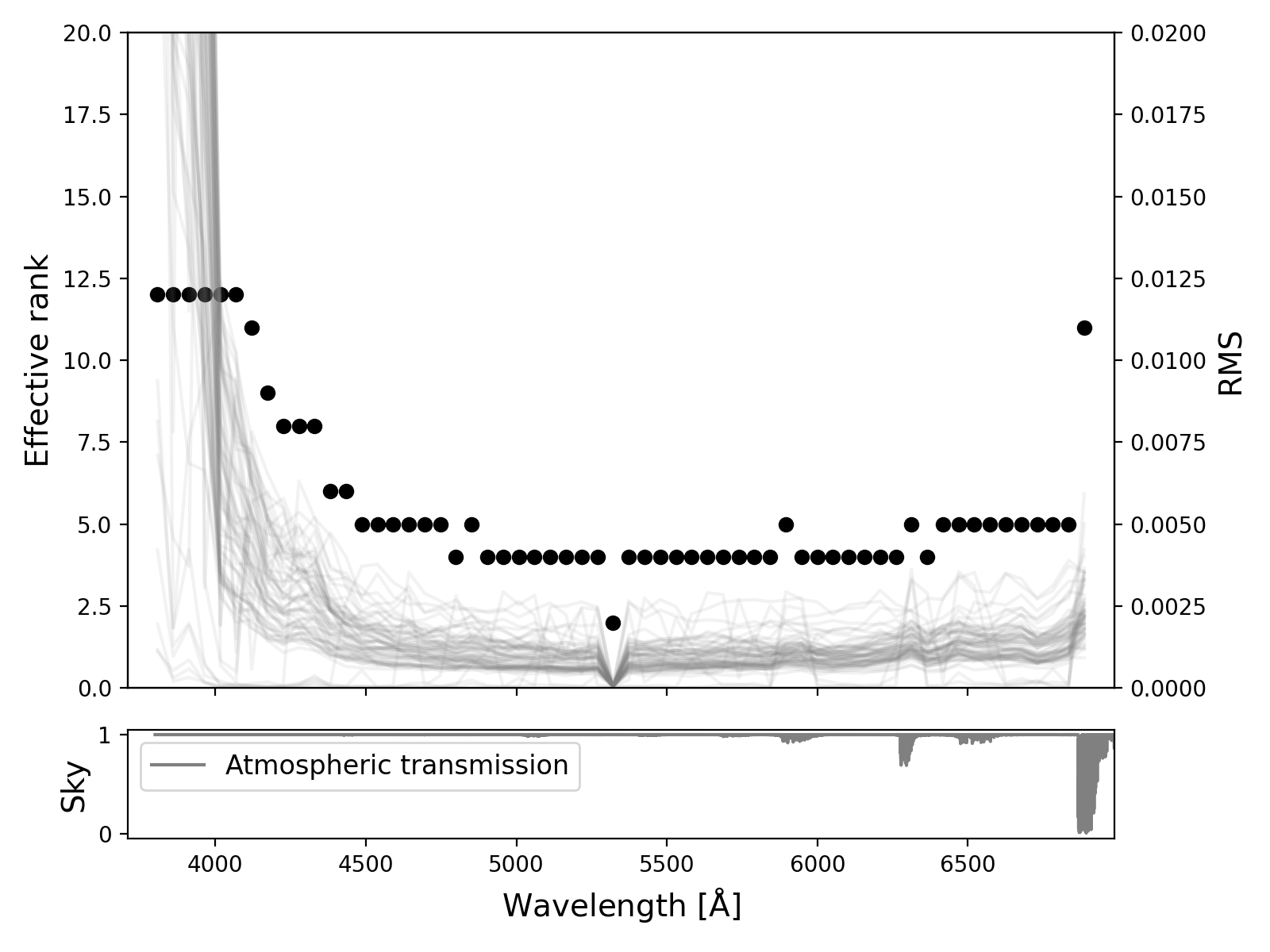}
    \caption{Upper panel: `Auto' SVD detrending for HARPS data. Black dots: Effective rank, Grey lines, RMS of the residual arrays as a function of wavelength. The gap between the two detectors is well visible at about 530\,nm. Lower panel: Atmospheric transmission. Wavelength areas with large RMS in the upper panel, match well with strong telluric lines, which are less correlated in the primary's rest frame.}
    \label{fig:eff_rank_HARPS}
\end{figure*}

\subsection{Generation of SVD side-lobes}

\begin{figure*}
	
        \includegraphics[width=0.6\linewidth]{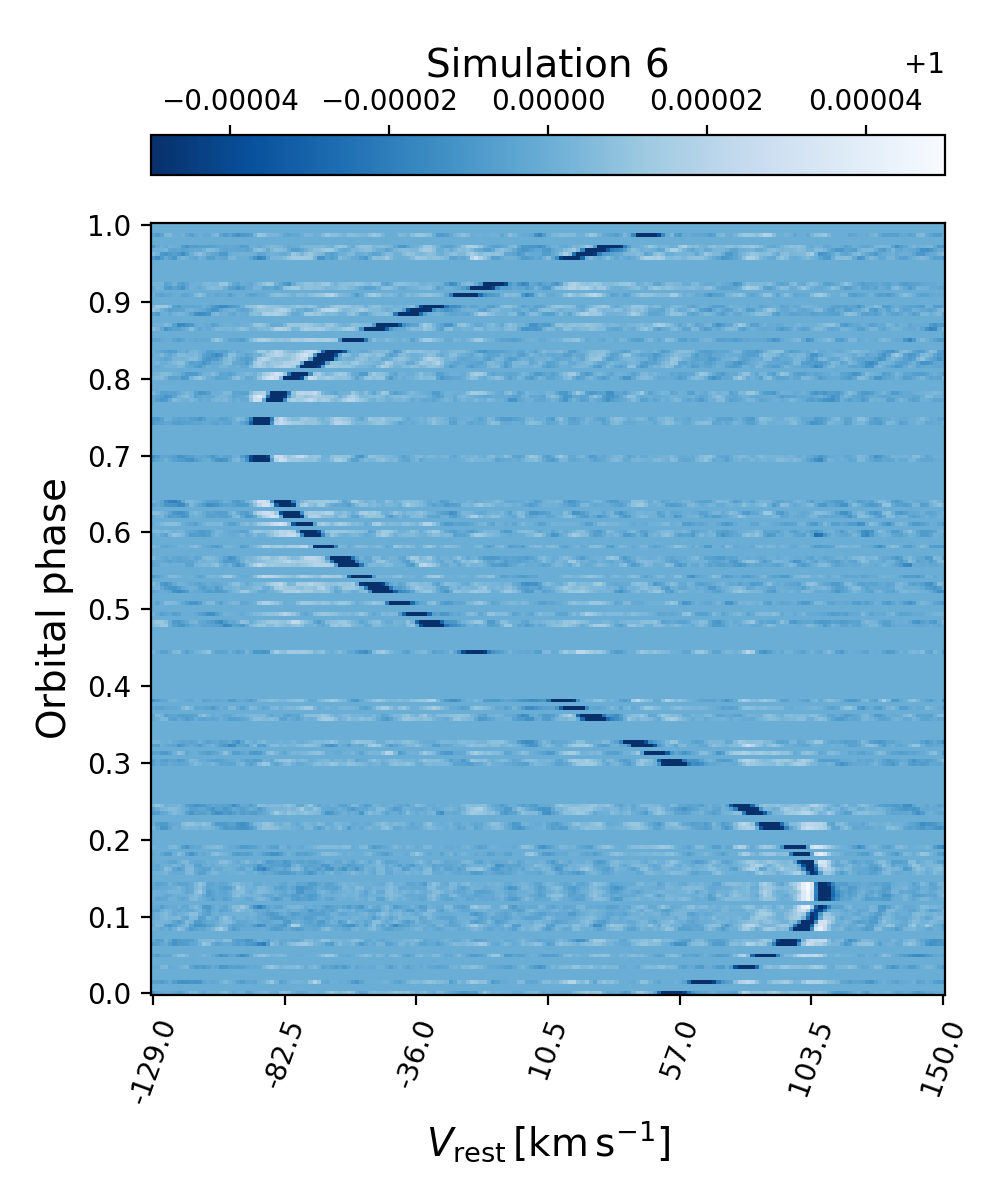}
        
    \caption{Cross-correlation functions for simulation 6 after SVD auto detrending in the rest frame of the primary. The CCF signal of the simulated secondary, as well as the wiggles of the CCF are clearly visible, following the secondary's orbital motion. The SNR of the secondary is increased by two orders of magnitude, compared to ESPRESSO data, making white structures visible. These represent residuals of the SVD detrending, forming artificial side-lobes in the average line profile.}
    \label{fig:A_side_lobes}
\end{figure*}

\subsection{SVD degradation for simulated data}

	
        

\begin{figure*}
        \includegraphics[width=0.48\linewidth]{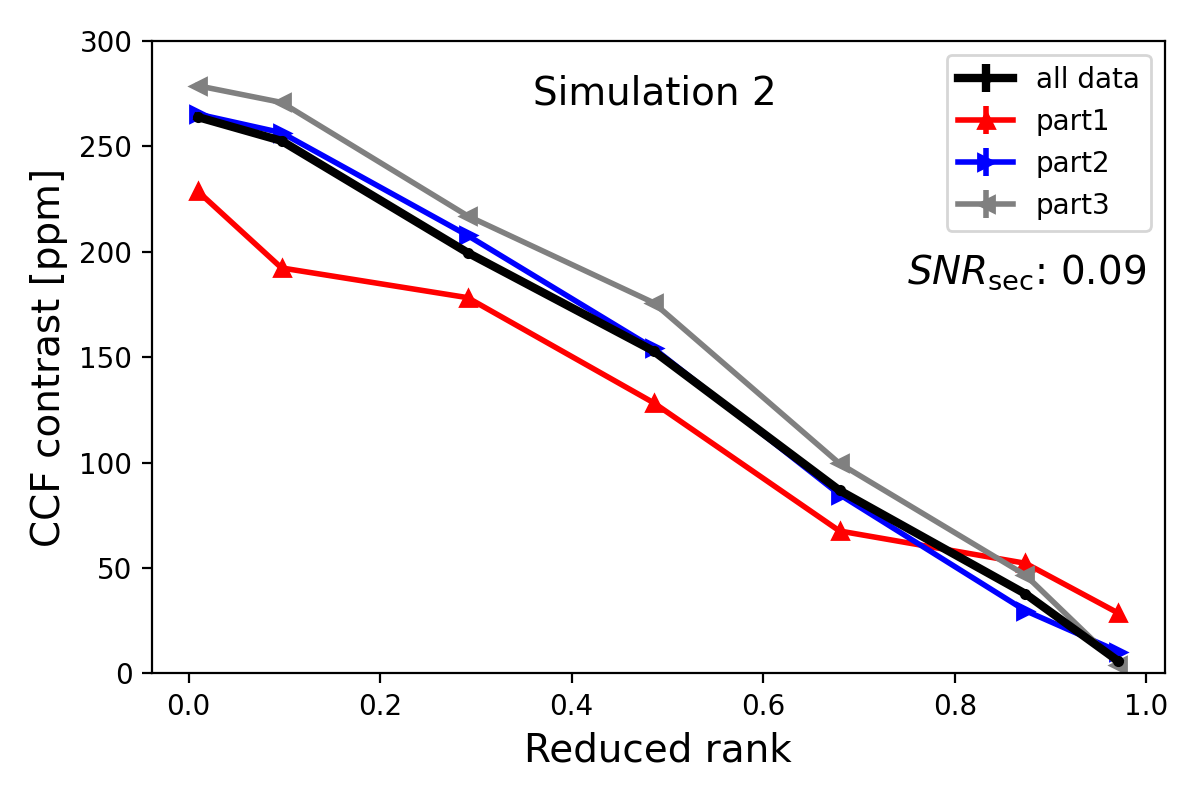}
        \includegraphics[width=0.48\linewidth]{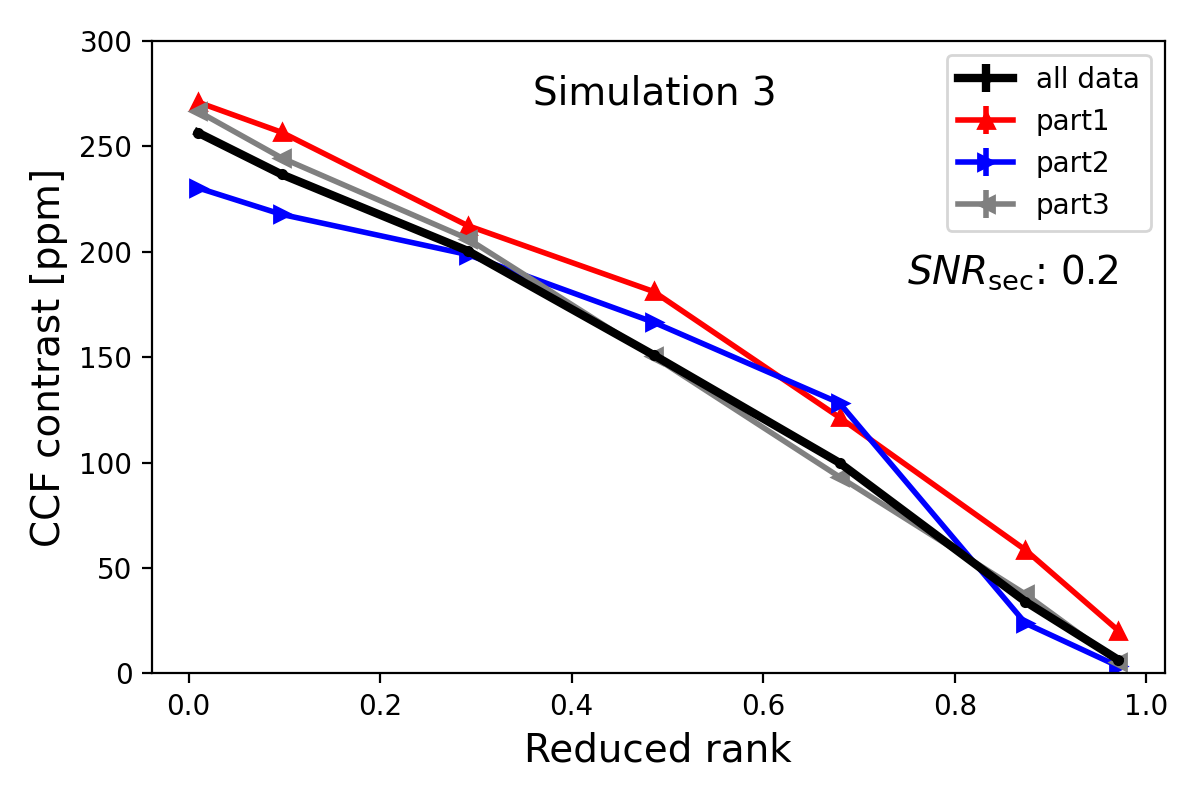}
        \includegraphics[width=0.48\linewidth]{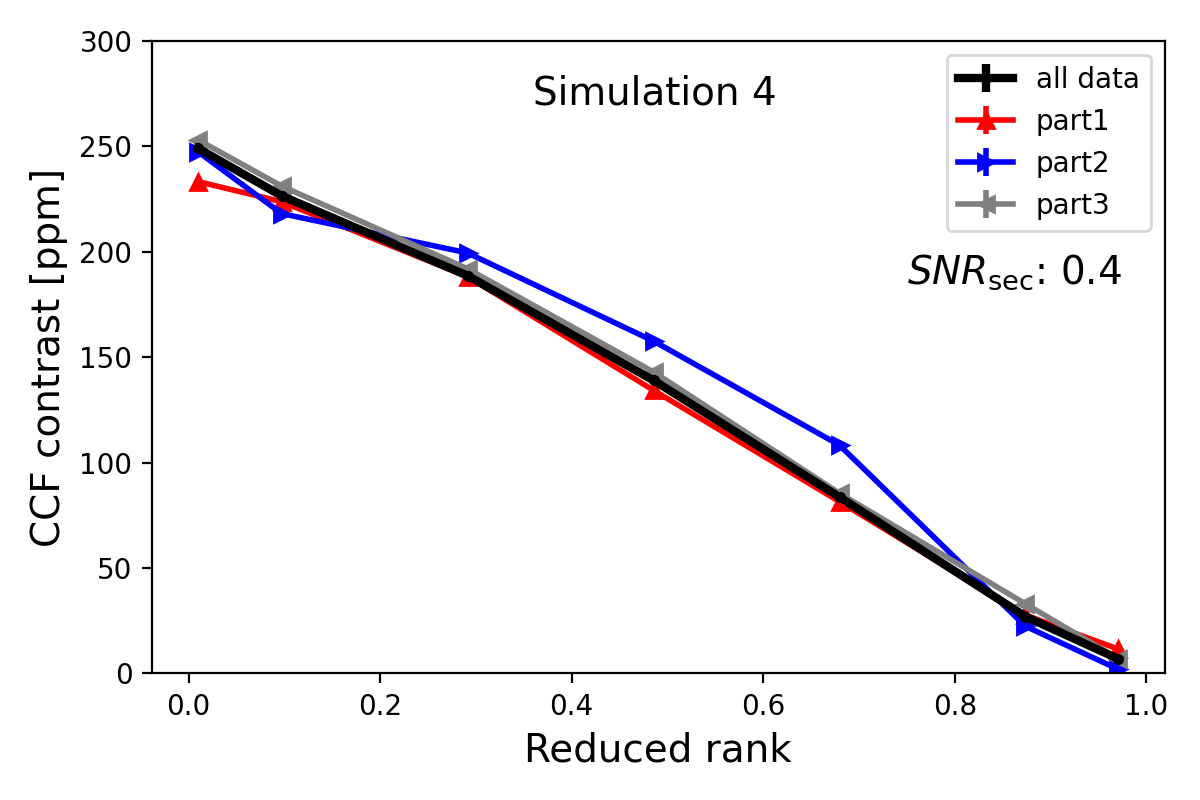}
        \includegraphics[width=0.48\linewidth]{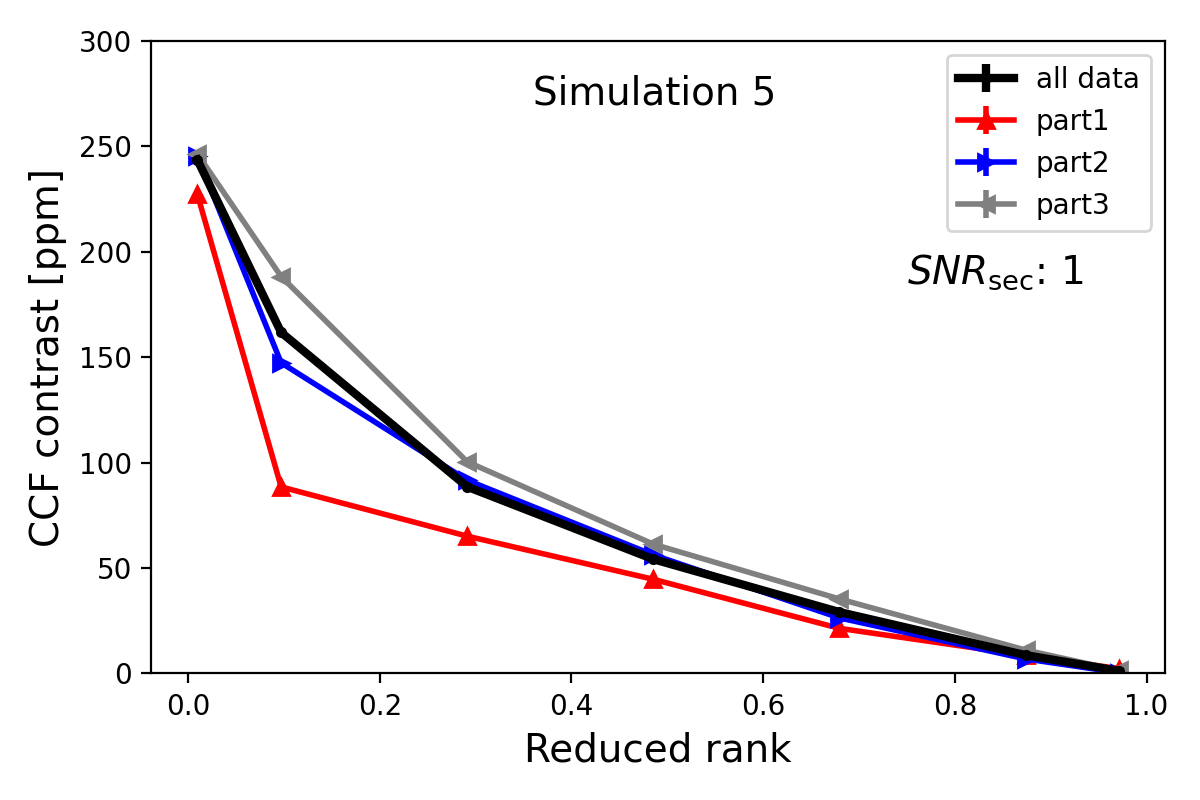}
        \includegraphics[width=0.48\linewidth]{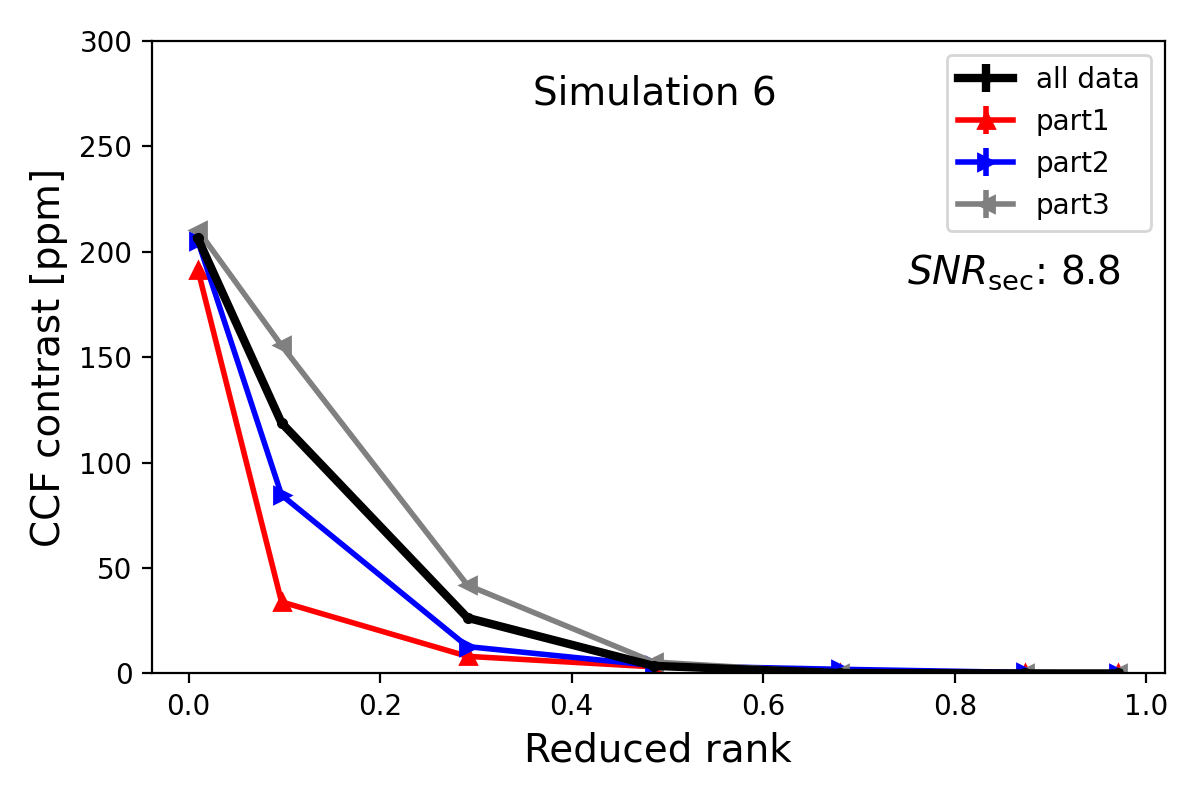}
        \includegraphics[width=0.48\linewidth]{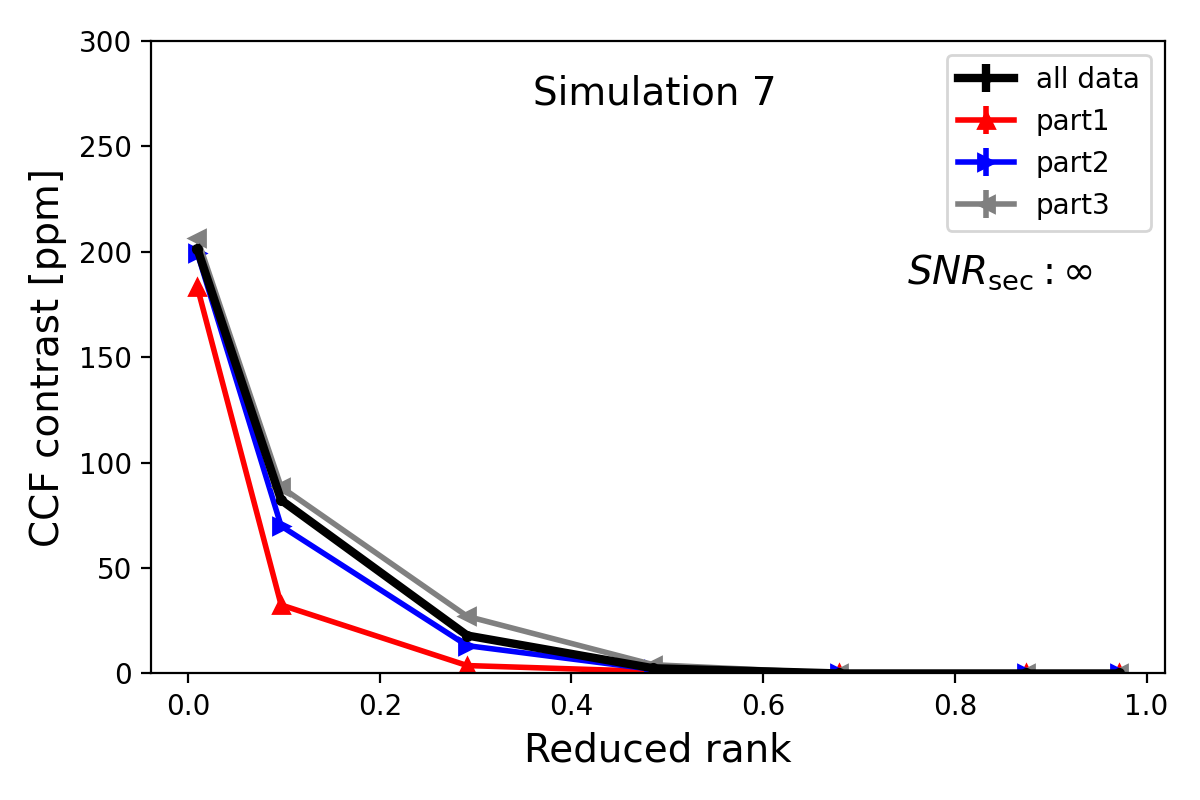}
        
    \caption{SVD degradation for modelled data versus reduced rank ($k/R$). Panels show simulations for different SNR of the secondary (per resolution element of each individual spectrum). Each panel shows the signal degradation for a selection of measurements according to their alignment in the primary's rest frame. Red: aligned with several other measurements (part1), Blue: aligned with one other measurement (part2) , Grey: not aligned (part3), Black: full sample. The degradation is approximately linear and not phase-dependent for secondary SNR$\ll 1$, but highly phase-dependent for secondary SNR$\gtrsim 1$.}
    \label{fig:A_erosion}
\end{figure*}

\section{Uncertainties from orbit parameters}

\begin{figure*}
	
        \includegraphics[width=\linewidth]{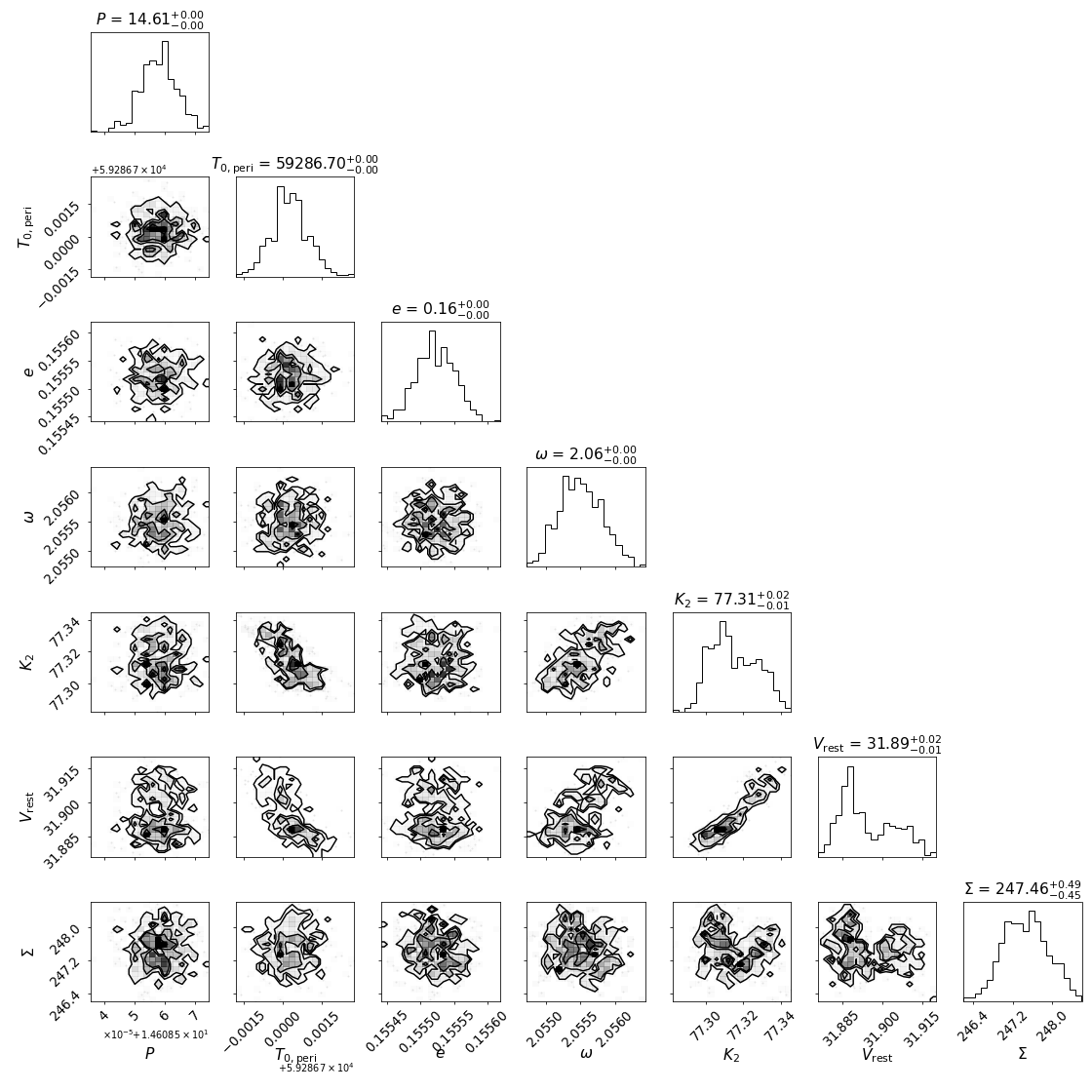}
    \caption{Dependence of fitted parameters from orbital uncertainties.}
    \label{fig:A_orbit}
\end{figure*}

\bsp	
\label{lastpage}

\end{document}